\documentclass[prb,aps,twocolumn,floats,floatfix,showpacs]{revtex4}
\usepackage{graphicx}
\usepackage{times}

\newcommand{\nin}{\noindent}

\newcommand{\be}{\begin{equation}}
\newcommand{\ee}{\end{equation}}
\newcommand{\bea}{\begin{eqnarray}}
\newcommand{\eea}{\end{eqnarray}}

\newcommand{\lb}{\left[}
\newcommand{\rb}{\right]}
\newcommand{\lp}{\left(}
\newcommand{\rp}{\right)}
\newcommand{\lf}{\left\{}
\newcommand{\rf}{\right\}}

\renewcommand{\H}{{\cal H}}
\renewcommand{\L}{{\cal L}}

\newcommand{\F}{{\cal F}}

\newcommand{\lsaw}{\lambda_{\rm ext}}
\newcommand{\ksaw}{k_{\rm ext}}

\newcommand{\ts}[1]{\textstyle{#1}}
\newcommand{\ds}[1]{\displaystyle{#1}}

\newcommand{\lch}{{l_{\rm ch}}}
\newcommand{\lfl}{{l_{\rm fl}}}
\newcommand{\lwf}{{l_{\rm wf}}}
\newcommand{\mtot}{m_{\rm tot}}

\begin{document}
\title{Electron properties of carbon nanotubes in a periodic potential}
\author{Dmitry S. Novikov}
\email{dima@alum.mit.edu}
\affiliation{Department of Electrical Engineering and Department of Physics,
Princeton University, Princeton, NJ 08544}
\date{\today}


\begin{abstract}

\nin
A periodic potential applied to a nanotube is shown to lock electrons 
into incompressible states that can form a devil's staircase. 
Electron interactions result in spectral gaps
when the electron density (relative to a half-filled Carbon $\pi-$band)
is a rational number per potential period, 
in contrast to the single-particle case  
where only the integer-density gaps are allowed. 
When electrons are weakly bound to the potential, 
incompressible states arise due to Bragg diffraction in 
the Luttinger liquid.
Charge gaps are enhanced due to quantum fluctuations, 
whereas neutral excitations are governed by an
effective SU(4)$\simeq$O(6) Gross-Neveu Lagrangian.
In the opposite limit of the tightly bound electrons, 
effects of exchange are unimportant,
and the system behaves as a single fermion mode that represents
a Wigner crystal pinned by the external potential, with the
gaps dominated by the Coulomb repulsion.
The phase diagram is drawn using the effective 
spinless Dirac Hamiltonian derived in this limit.
Incompressible states can be detected in the adiabatic transport setup
realized by a slowly moving potential wave,
with electron interactions providing the possibility of  
pumping of a fraction of an electron  
per cycle (equivalently, in pumping at a fraction of the base frequency).

\end{abstract}

\maketitle

\section{Introduction}
\label{sec:intro}

Since their discovery,\cite{Iijima} carbon nanotubes (NTs) 
remain in focus of both basic and applied research.
\cite{Dresselhaus,AvourisDresselhaus} 
Besides their important technological potential,
\cite{McEuen'98,Dekker'99} 
nanotubes are a testing ground for
novel physical phenomena involving strong electron interactions.
In particular, 
they are believed to be perfect systems to study 
Tomonaga-Luttinger liquid effects.
\cite{Krotov'97,Kane'97,Egger'97,Odintsov'99,Levitov'01,Stone}
Experimentally, effects of electron-electron interactions in nanotubes
have been observed
in the Coulomb blockade peaks in transport,
\cite{Bockrath'97,Tans'98} 
in the power law temperature and bias dependence of the
tunneling conductance,\cite{Bockrath'99,Yao'99,Nygard} 
and in the power law dependence
of the angle-integrated photoemission spectra.\cite{PE}

Nanotubes are rich systems to study electron correlations
for a number of reasons.
Indeed, their one-dimensional (1d) character increases effects of interactions;
spin and Brillouin zone degeneracy\cite{Dresselhaus} result in 
the presence of the four polarizations of Dirac fermions 
which can allow one to probe SU(4) spin excitations;
\cite{Levitov'01,SU4Kondo-exp,SU4Kondo-th} 
exceptional chemical and mechanical NT properties result in very low disorder; 
finally, diverse methods of nanotube synthesis, 
a variety of available nanotube chiralities and of the ways of coupling 
to the nanotube electron system allow one to explore a wide region of 
parameter space.

In the present work we propose to employ
the coupling of an external periodic potential
to the nanotube electronic system as a probe of both Tomonaga-Luttinger 
correlations, and of the 1d Wigner crystallization, via commensurability effects.
We focus on the electron properties of single-wall NTs
in a periodic potential whose period $\lsaw$ is much 
greater than the NT radius $a$, $\lsaw\gg a$.
Such a potential can be realized using optical methods, by gating, or by 
an acoustic field. In all of these cases, the realistic period $\lsaw$ is of 
the order 0.1--1\,$\mu$m. 
As shown below, effects of the Tomonaga-Luttinger correlations 
on the Bragg diffraction of electrons, realization of the 
SU(4) spin excitations, as well as pinning of the 
Wigner crystal can be demonstrated in such a
setup depending on the applied potential and on the NT parameters.

In particular, below we will identify {incompressible electron states}, 
characterized by excitation gaps, that can arise
when the average NT electron number density $\bar\rho$ 
(counted from half-filling) is
{commensurate} with the period $\lsaw$ of the external potential:
\be \label{rho-alpha}
\bar\rho 
= {\mtot \over \lsaw} \,,  \quad \mtot = 4m \,.
\ee
In Eq.~(\ref{rho-alpha}), $m$ is the number of fermions of each 
of the four polarizations (herein called ``flavors'') per period.

At what density values $m$ can the spectral gaps open?
The single-particle treatment\cite{Talyanskii'01}
maps the problem onto that of the Bloch electron, resulting 
in the spectrum of {minibands} separated by 
{minigaps} as a consequence of the Bragg diffraction on the external 
potential. Filling up an integer number $m$ of minibands corresponds
to adding $m$ electrons of each flavor per ``unit cell'' (the period $\lsaw$). 
Thus, minigaps open up when the density (\ref{rho-alpha}) is {\it integer}:  
$m=0, \, \pm 1, \,\pm 2, \, ...$ .\cite{Talyanskii'01}
In other words, the wave nature of the Bloch electron 
leads to commensurate states with integer density.

\begin{figure}
(a)~~~~~~~~~~~~~~~~~~~~~~~~~~~~~~~~~~~~~~~~~~~~~~~~~~~~~~~~~~~~~~~~~~~~\\
\includegraphics[width=3.5in]{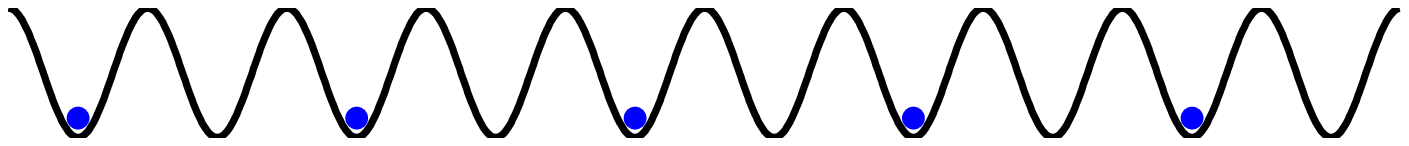}
(b)~~~~~~~~~~~~~~~~~~~~~~~~~~~~~~~~~~~~~~~~~~~~~~~~~~~~~~~~~~~~~~~~~~~~\\
\includegraphics[width=3.5in]{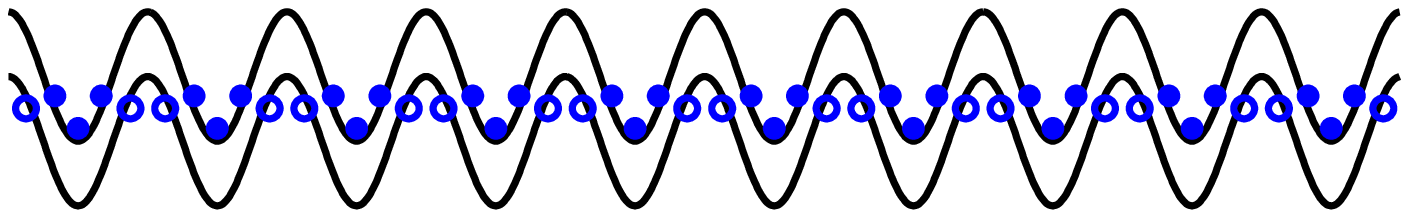}
(c)~~~~~~~~~~~~~~~~~~~~~~~~~~~~~~~~~~~~~~~~~~~~~~~~~~~~~~~~~~~~~~~~~~~~\\
\includegraphics[width=3.5in]{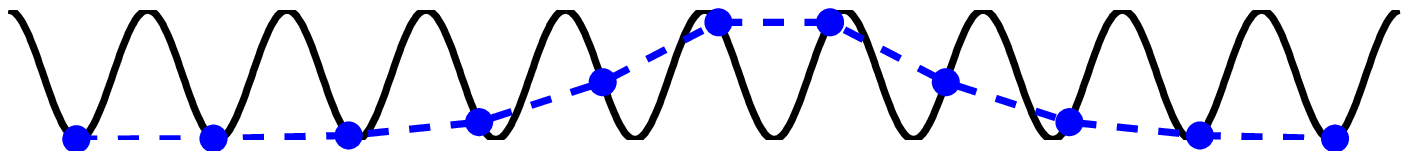}
(d)~~~~~~~~~~~~~~~~~~~~~~~~~~~~~~~~~~~~~~~~~~~~~~~~~~~~~~~~~~~~~~~~~~~~\\
\includegraphics[width=3.5in]{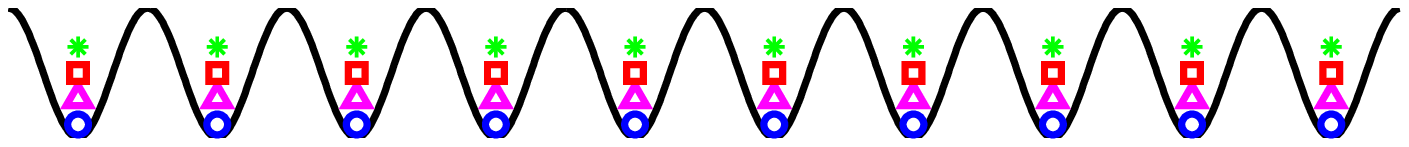}
(e)~~~~~~~~~~~~~~~~~~~~~~~~~~~~~~~~~~~~~~~~~~~~~~~~~~~~~~~~~~~~~~~~~~~~\\
\includegraphics[width=3.5in]{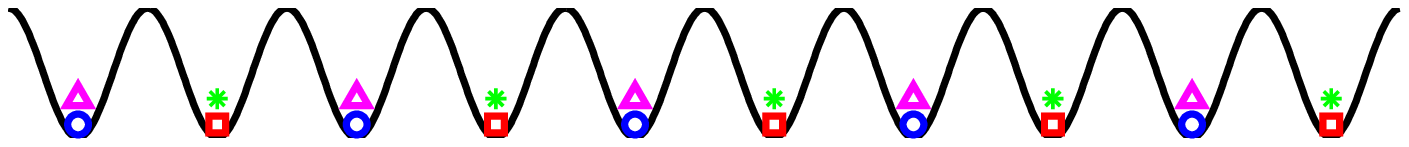}
\caption[]{(Color online)
Incompressible electron states (schematic).
(a) The $(1/2,0)$ state with $\mtot=1/2$.
(b) The $(3,2)$ state with $\mtot=1$.
Large potential amplitude breaks the ``Dirac vacuum'' 
placing holes (open circles) into the potential maxima.
(c) An example of the {\it phase soliton excitation} 
over the $(1,0)$ state. The excitation gap is given by the phase soliton energy. 
(d) The $m=1$ incompressible state allowed by the Bloch theory of 
noninteracting nanotube electrons. Four kinds of labels 
mark the NT fermions of the four flavors.   
(e) Semiclassical picture of the simplest interaction-induced $m=1/2$ state. 
Fermions of the same flavor avoid each other due to the Pauli principle. 
}
\label{fig:soliton-sketch}
\end{figure}

Our main result here is that electron interactions 
dramatically change the spectrum, adding 
incompressible states at {\it rational} densities $m=p/q$,
in which commensurability is induced by interactions.
\cite{NT-devil}
In a fractional$-m$ state, the NT electron system is 
locked by the external potential into a $q\lsaw-$periodic 
commensurate configuration (such as the ones schematically represented 
in Fig.~\ref{fig:soliton-sketch}).  
Naturally, the states with the lower denominator $q$ are
more pronounced. Realistically, due to finite NT length and 
temperature, only a few states with small enough $q$
can be detected.
However, these fractional-$m$ states are important as the corresponding
minigaps are interaction-induced (and vanish in the noninteracting limit).
Measurement of such minigaps can provide a direct probe 
of interactions between electrons.

From a practical standpoint, measurement of the gaps can identify
the incompressible states characterized by
quantum coherence on a macroscopic scale. 
A challenging experimental proposal\cite{Talyanskii'01} is to 
realize the Thouless pump\cite{Thouless}
by taking advantage of the semi-metallic NT dispersion.
In such a setup, a {\it quantized current} is predicted to arise whenever 
the chemical potential is inside the minigap created by the adiabatically
slowly moving potential wave. 
As our approach suggests, due to electron interactions, 
commensurability (\ref{rho-alpha})
will result in additional current plateaus\cite{NT-devil}
corresponding to pumping on average of a {fraction of unit charge per cycle}. 
Equivalently, electron interactions could allow one to realize 
a novel effect of adiabatic pumping of charge 
at the fraction of the base frequency of the potential modulation.

In connection with the adiabatic current quantization, we note a parallel between
our system and the quantum Hall effect. In both cases, commensurability 
with external electromagnetic field yields incompressible states, spectral 
gaps, and quantization in transport.
Moreover, it is the electron interactions that result in 
the incompressible states at fractional filling factors
(commensuration due to interactions) in addition to the integer plateaus 
(commensuration due to wave nature of electrons),
both causing non-dissipative transport.\cite{Kane-Lubensky}
This similarity extends onto the disorder stabilizing
plateaus at lower denominators.

From a theoretical viewpoint, the excitation spectrum 
is linked to the general theory of commensurate-incommensurate transitions.
\cite{Frenkel,FvdM,Dzyal,Pokrovsky,Bak}
In this approach, 
an excitation over the commensurate state (an incommensuration) is
represented by a {\it phase soliton} 
[illustrated in Fig.~\ref{fig:soliton-sketch}(c)], 
whose energy gives the corresponding spectral gap.
In the present work the phase soliton method is generalized onto the 
case of strongly interacting massive Dirac fermions of multiple polarizations.

Technically, the problem at hand requires non-perturbative treatment of 
electron interactions at low density ($\bar\rho \lsaw = \mtot \sim 1$),
i.e. at the bottom of the band, taking into account the effects
of the {curvature} of the electronic dispersion.
This is hard to achieve in the conventional bosonization scheme
which rests on the linearized dispersion and
describes hydrodynamic density modes extended over the whole system.
Indeed, the curvature of the dispersion is a 
``dangerously irrelevant perturbation'' that causes ultraviolet divergencies.
However, it is the curvature of the dispersion that can provide a length scale 
to an otherwise scale-invariant Gaussian bosonic action.
Not surprisingly, phenomena in which the physics on the scale of Fermi wavelength 
is important, such as the Coulomb drag between quantum wires,\cite{Drag}
often rely on the curvature and, hence, are difficult to treat.

Nanotubes provide a crucial theoretical advantage: Even {\it massive} 
interacting Dirac fermions can be bosonized, by virtue of 
the massive Thirring --- sine-Gordon duality,\cite{Coleman,Haldane}
with the curvature of the dispersion controlled by the Dirac 
mass (NT gap $\Delta_0$ at half-filling). In the bosonic language, 
the gap controls the strength of the nonlinear term in the bosonic action.
Moreover, the NT Hamiltonian is SU(4) symmetric in the forward scattering 
approximation. This symmetry simplifies the treatment since 
in the problem of the four coupled modes, 
only two different length scales appear: The charge scale, $\lch$
(screening length for the Coulomb interaction), and the flavor scale, 
$\lfl \ll \lch$, with a meaning of a size of the electronic wave function
represented by a composite sine-Gordon soliton of the charge and flavor modes.
\cite{Levitov'01}

External periodic potential locks electrons
into incompressible states at commensurate densities.
As a semiclassical illustration of this phenomenon,
in Fig.~\ref{fig:soliton-sketch}(d,e) the centers of solitons that represent 
electrons of the different flavors are marked by the four different symbols.
In accord with the above, such a locking is only possible when 
the NT gap $\Delta_0\neq 0$,
since the dispersion curvature gives rise to the finite soliton sizes 
$\lch, \lfl \propto \Delta_0^{-\gamma/2}$ with $\gamma>0$. 
For strictly linear dispersion 
(as in the armchair tube), both of these scales are infinite, 
and the minigaps do {\it not} open, as 
it is impossible to pin a system that has an infinite length scale.\cite{FSA}

In general, both Bragg diffraction and the Coulomb interaction contribute to
the excitation gaps. While the diffraction is a dominant factor
for the incompressible states when the system is
controlled by the Luttinger-liquid fixed point, the Coulomb interaction 
plays a major role when the electron system is in the Wigner-crystal regime.
The external potential, by bringing the additional length scale $\lsaw$
to the problem, naturally distinguishes between the two sides of the 
Luttinger-liquid -- Wigner-crystal crossover.
Technically, this distinction occurs on the level of a saddle point 
of the nonlinear bosonic action.

In a recent Letter, \cite{NT-devil} we showed that, 
in the semiclassical (strong coupling) limit, when
electrons are nearly pointlike ($\lfl \ll \lsaw$), 
the long range Coulomb interaction
leads to the devil's staircase of commensurate states representing 
the pinned Wigner crystal.
Gaps open at rational densities $\mtot$, and Coulomb interaction
sets the gap energy scale.
An example of such a state with $\mtot=1/2$ is shown schematically in
Fig.~\ref{fig:soliton-sketch}(a). 
In this work we will study the phase diagram in this limit in detail.
Remarkably, 
the {Dirac} character of nanotube electrons  
brings about a set of incompressible states
in which the ``Dirac vacuum'' is broken when the potential amplitude exceeds 
the gap at half-filling. In this case, {physically different} 
incompressible states can correspond to the  {same} total density 
(\ref{rho-alpha}) with
\be \label{ne-nh}
4m=\mtot=n_e-n_h \,.
\ee 
To further characterize these states we specify the pair of 
numbers $(n_e, n_h)$ of {electrons} and {holes} in the potential 
minima and maxima correspondingly. An example of the $(3,2)$ state with 
$\mtot=1$ is schematically shown in Fig.~\ref{fig:soliton-sketch}(b).

The opposite limit of the weakly coupled electrons ($\lfl \gg \lsaw$)
is connected to the gaps opening due to the Bragg diffraction. 
In this case electrons are delocalized over many periods, 
and gaps occupy a small part of the spectrum.
Exchange is important, and adds to the cost of an excitation in which
an extra electron of a particular flavor is added to the commensurate configuration.
Below we will specifically focus on the lowest-denominator fractional 
state $m=1/2$, for which we find the effective action that describes 
excitations over the commensurate state, and obtain the charge and the 
SU(4)--flavor excitation gaps.

This paper is organized as follows.
In Section~\ref{sec:model} we introduce the model Hamiltonian. 
In Section~\ref{sec:nonint}, as a warm-up, we consider the noninteracting problem.
In Section~\ref{sec:boson} the many-body Hamiltonian is bosonized. 

Section~\ref{sec:int} is central, as we introduce the phase soliton 
method for the bosonized NT electrons.
In Section~\ref{sec:weak} we consider the Bragg diffraction limit, 
whereas in Section~\ref{sec:strong} we focus on the pinned Wigner crystal.
In Section~\ref{sec:weak-tunneling} we outline the phase diagram 
in the semiclassical (Wigner crystal) limit.
In Section~\ref{sec:current} we discuss experimental ways
to detect incompressible states.

\section{The Model}
\label{sec:model}

Near half-filling, the nanotube electron system in 
the forward scattering approximation\cite{Egger'97,Kane'97}
is described in terms of the four ($2_{\rm spin}\times 2_{\rm valley}$) 
Dirac fermion flavors, with the 
following second-quantized Hamiltonian, 
\be \label{H}
\H=\H_0 + \H_{\rm bs} + \H_{\rm ext} \,.
\ee
The first term $\H_0$ is the massless Dirac Hamiltonian
that includes Coulomb repulsion:
\be \label{H0}
\H_0 = -i\hbar v\int 
\sum_{\alpha = 1}^4\psi^+_{\alpha}\sigma_3\partial_x\psi_{\alpha} dx 
+ \frac{1}{2}\sum_k\rho_kV(k)\rho_{-k}\,. 
\ee
Here 
\be \label{psi-alpha}
\psi_{\alpha} = \lp \matrix{\psi^R_{\alpha} \cr \psi^L_{\alpha}} \rp \,, 
\quad 1\leq \alpha \leq 4
\ee
is a two component Weyl spinor of the flavor $\alpha$, 
$x$ is the coordinate along the tube, 
$\sigma_{1,3}$ are the Pauli matrices, 
and $v\approx 8\cdot 10^7$~cm/s is the NT Fermi velocity. 
In the total number density $\rho_k=\int\! dx\, e^{-ikx}\rho(x)$, with
\be \label{rho}
\rho(x) \simeq \sum_{\alpha = 1}^4\psi^+_{\alpha}(x)\psi_{\alpha}(x) \,,
\ee
we neglected the strongly oscillating components $\sim e^{\pm 2iK_{\rm BZ}x}$, with 
$K_{\rm BZ}\sim 1/a_{\rm cc}$ 
defining the position of the Dirac points in the Brillouin zone of graphene
[$a_{cc}=0.143$~nm being the length of the Carbon bond].
For a nanotube of radius $a$ 
placed on a substrate with the dielectric constant $\varepsilon$, 
the 1d Coulomb interaction 
\be \label{V}
V(k) = \textstyle{\frac{2}{\varepsilon+1} V_0(k)} \,,
\ee
where
\be \label{V0}
V_0(k) \simeq e^2\ln\lb 1+(ka)^{-2}\rb \,.
\ee

The essence of the forward scattering approximation employed above is 
that even in the presence of electron interactions,
the two Dirac points at the corners of the Brillouin zone
remain decoupled. 
Since the long range potential (\ref{V}) does not discriminate between the 
Carbon sublattices, scattering amplitudes 
for the fermions of same and opposite chiralities at each Dirac point are equal.
\cite{Egger'97,Kane'97} 
Hence in the Hamiltonian (\ref{H0}) we include
only the part of electron interaction that   
involves the smooth part (\ref{rho}) of the electron density.
Furthermore, in the Hamiltonian (\ref{H0})
the backscattering and Umklapp processes
between the Dirac points are discarded. 
These processes are parametrically reduced by $a_{cc}/a \lesssim 0.1$, 
while the Umklapp amplitude is also numerically small.\cite{Odintsov'99}

The second term in the Hamiltonian (\ref{H}) describes the backscattering
between the left and right moving fermions {\it within} each Dirac point:
\be \label{Hbs}
\H_{\rm bs}  =  
\Delta_0 \int 
{\textstyle \sum_{\alpha = 1}^4} \psi^+_{\alpha}\sigma_1\psi_{\alpha} dx \,.
\ee
This backscattering is different from the usual ``$V(2k_F)$'' term in
quantum wires, as the interaction-induced gaps are undetectably 
small in metallic NTs.
Instead, we rely on the curvature of the dispersion arising from the bare  
gap $\Delta_0$ at half-filling. Remarkably, this gap can 
greatly vary depending on NT chirality or external fields.\cite{Dresselhaus}
In particular, 
semiconducting nanotubes have a large gap 
\be \label{Delta0-semicond}
\Delta_0^{\rm (semicond)}={\hbar v\over 3a} \simeq 0.18~{\rm eV}/a_{\rm [nm]} 
\ee
that scales inversely with the radius $a$.
Metallic NTs can be of the two kinds.
There are truly metallic, or the so-called ``armchair'' NTs 
which have a zero gap at half-filling, $\Delta_0=0$.
However, a gap 
$\Delta_0 \ll D$
that is small compared to the 1d bandwidth 
\be \label{D}
D = {\hbar v \over a} = 0.53~{\rm eV}/a_{\rm [nm]} 
\ee
can appear due to the curvature of the
2d graphene sheet in the nominally metallic tube. 
This gap is inversely proportional to the square of the NT radius,
and is numerically given by
\be
\Delta_0^{\rm (semimet)} 
\approx 10~{\rm meV}\cdot |\cos 3\Theta_{\rm ch}|/a_{\rm [nm]}^2
\ee
as a function of the NT {chiral angle} $\Theta_{\rm ch}$.
\cite{zigzag,exp-curvature}  
Even smaller $\Delta_0$ can be induced in a strictly metallic
``armchair'' ($\Theta_{\rm ch}=\pi/6$) tube by applying
magnetic field parallel to the NT axis. \cite{B-paral,exp-B-parallel}

Finally, interaction with the external periodic potential 
is represented by 
\be \label{Hext}
\H_{\rm ext} = \int \! dx \, \rho \, U(x) \,.
\ee
In the present work, for the purpose of simpler algebra, we consider 
the potential of the form
\be
\label{SAW}
U(x) = A \cos \ksaw x \,, \quad \ksaw={2\pi / \lsaw} \,.
\ee
Qualitatively, the results of this work will be valid for any periodic potential
realized by the means described in the Introduction.
As the values of the period $\lsaw$ are in the 
$0.1 - 1\,\mu$m range,  the separation of scales $\ksaw \ll 1/a \ll 1/a_{cc}$
rules out a possibility of coupling between the Dirac points via 
the potential (\ref{SAW}), in full compatibility
with the forward scattering approximation.

In general, our main conclusion about the interaction-induced
incompressible states at fractional densities that open up in addition to 
the integer-density gaps is valid for any value of the NT gap $\Delta_0$.
Quantitatively, our present analysis that will be based on the bosonization 
demands the separation of scales $\Delta_0 \ll D$, 
practically requiring the semimetallic tubes.

The Hamiltonian (\ref{H}) written in the forward scattering approximation
is SU(4) invariant with respect to rotations in the space of the 
fermion flavors $\psi_{\alpha}$, with both the Coulomb interaction
and the coupling to the smooth external potential (\ref{SAW})
preserving this symmetry.

\section{Noninteracting Electrons}
\label{sec:nonint}

In the absence of interactions, the contributions of the four fermion flavors
factorize. Here we consider the single electron spectrum 
[cf. Eqs.~(\ref{H0}), (\ref{Hbs}), and (\ref{Hext})] 
\be \label{eigenvalueproblem}
\lf \H_D+U(x) \rf \psi = \epsilon \psi \,,
\ee
where $\H_D$ for fermions of each flavor is given by 
\be \label{H-Dirac}
\H_{D} = -i \hbar v \, \sigma_3 \partial_x + \Delta_0 \, \sigma_1  \,.
\ee 
This spectrum has been briefly analyzed 
in Ref.~\onlinecite{Talyanskii'01}. 
Below we will study it in detail emphasizing 
the distinction between the two opposite regimes of coupling,
tight-binding {\it vs.} nearly-free electrons.
It is convenient to perform a gauge transformation 
\be \label{psi-gauge}
\psi'(x) = e^{-i \bar A \sigma_3 \sin \ksaw x}\, \psi(x) \,,
\ee
where  the dimensionless  amplitude
\be \label{bar-A}
\bar A = {A/ \epsilon_0} \,.
\ee
The kinetic energy scale 
\be \label{epsilon0}
\epsilon_0 \equiv \hbar \ksaw v  = hv/\lsaw = 3.3~{\rm meV}/{\lsaw}_{\rm \, [\mu m]}
\ee
is the Dirac level spacing in each potential minimum.

After the gauge transformation (\ref{psi-gauge}) the 
Hamiltonian $\H_D+U(x)$ becomes
\be \label{H-nonint-gauged}
\H' = -i\hbar v \, \partial_x \sigma_3 
+ \Delta_0 \sigma_1 e^{-2i\bar A \sigma_3 \sin \ksaw x} \,.
\ee
Remarkably, due to the Dirac character of the problem, in the Hamiltonian 
(\ref{H-nonint-gauged})
the relative importance of the potential energy (the second term)
is governed by the value of the Dirac gap $\Delta_0$, rather than
the potential amplitude $A$,
whereas the kinetic energy (the first term) is ${\cal O}(\epsilon_0)$.
[Note that in Eq.~(\ref{H-nonint-gauged}) 
we discarded the Schwinger anomaly term $\propto \int \! dx \, U^2(x)$ 
since its effect\cite{NT-anom} of adding a constant energy shift 
is not important here.]

The coupling of the NT electron to the external potential can be either weak
or strong, depending on the gap value $\Delta_0$.
Consider the eigenvalue problem
\be \label{transfer-mat}
\H'\psi' = \epsilon \psi'
\ee
that is periodic in $\lsaw$. Its solutions are the spinor 
Bloch states $\psi'_p(x)=u_p(x) e^{ipx}$ with a quasimomentum $\hbar p$ 
taking values in the effective Brillouin zone defined by the potential period, 
$-\ksaw/2<p<\ksaw/2$.  Below we distinguish between the weak coupling limit 
(nearly free electrons), and the strong coupling (tight-binding) limit.

In the limit of {\it nearly free electrons}, 
\be \label{def-weak-nonint}
\psi'_p(x)=u_p(x) e^{ipx} \,, \quad 
u_p(x) \simeq {\rm const} ,
\ee
and the electron wave functions are close to plane waves.
Note that in the {\it massless} case of $\Delta_0=0$, $u_p \equiv {\rm const}$,
and the external potential has no effect since it is gauged away.
It is a manifestation of the fact that the scalar external potential
does not mix massless Dirac branches whose wave functions have a spinor structure.
This is also consistent with the above discussion that 
the potential energy scale is determined by the backscattering $\Delta_0$.

When the backscattering $\Delta_0\neq 0$, Bragg 
diffraction on the potential results in mixing between the left and right 
moving states of the Dirac spectrum at the values 
\be
p_m =\pm {m \over 2} \, \ksaw \,, \quad m=\pm 1, \, \pm 2, \, ... 
\ee
of the electron quasimomentum, 
opening minigaps in the single particle spectrum
at energies  $\epsilon_m=m\epsilon_0/2$. 
Perturbation theory in $\Delta_0$ yields 
the minigap values\cite{Talyanskii'01} $2\Delta_m^{(0)}$,  
\be
\label{minigaps-nonint}
\Delta_m^{(0)}(\bar A)=\Delta_0\, |J_m(2\bar A)| \,.
\ee
The superscript $(0)$ in Eq.~(\ref{minigaps-nonint})
relates to the noninteracting case.
Minigaps (\ref{minigaps-nonint}) oscillate as a function of the potential
amplitude $A$ and vanish for particular values of $A$ corresponding 
to zeroes of the Bessel functions $J_m$.

\begin{figure}[t]
\includegraphics[width=3.4in]{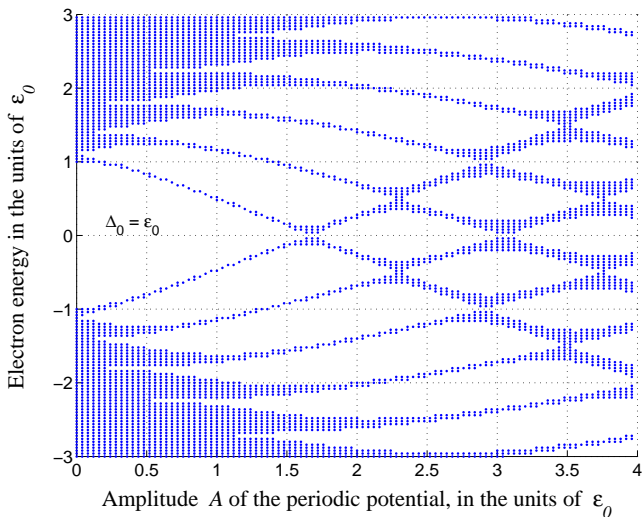}
\caption[]{(Color online)
Single electron spectrum 
in the potential (\ref{SAW})
for the case $\Delta_0=\epsilon_0$.
}
\label{fig:minibands}
\end{figure}

The opposite regime is the {\it strong coupling}, or the {\it tight-binding limit},
in which case the external potential localizes 
the semi-classical Dirac electrons (holes) in its minima (maxima),
tunneling between adjacent potential wells is exponentially suppressed,
and minigaps become larger than the subbands.
In this regime the wave function $u_p(x)$ becomes localized 
around the potential minimum, with its characteristic size 
\be \label{def-strong-nonint}
\lwf \ll  \lsaw \,.
\ee
The spectrum can be obtained by solving the problem (\ref{transfer-mat})
numerically via the transfer matrix.
For an illustration, in Fig.~\ref{fig:minibands} we plot the spectrum 
for the borderline case $\Delta_0 = \epsilon_0$ in which the potential 
and kinetic energy in the Hamiltonian (\ref{H-nonint-gauged}) are of the same order.
Similar to the nearly free electron limit, the 
energy gaps oscillate as a function of $A$ and vanish at certain 
amplitude values. The spectrum also has a characteristic Dirac 
$\epsilon\to -\epsilon$ symmetry.
Fig.~\ref{fig:minibands} suggests that even for 
moderately strong backscattering $\Delta_0=\epsilon_0$ 
minigaps can occupy most of the spectrum when $A\sim \epsilon_0$.

In Appendix~\ref{app:tb} we show that the limit (\ref{def-strong-nonint})
of exponentially suppressed interwell tunneling holds whenever
\be \label{tb-condition}
\lp \Delta_0\over \epsilon_0\rp^{-1} < {A\over \epsilon_0} < 
\lp \Delta_0\over \epsilon_0\rp^c \,, \quad \Delta_0 > \epsilon_0 \,,
\ee
where the exponent $c=3$ for tunneling from an energy level close to 
the bottom of the potential (or close to the top for holes), and 
$c=2$ for tunneling from a level far from the potential minimum or maximum.

Increasing the potential amplitude $A$ beyond $\Delta_0^c/\epsilon_0^{c-1}$
actually {\it enhances} the tunneling amplitude between the particle and 
hole continua. This is reflected in Fig.~\ref{fig:minibands} by 
wider minibands for $A > \epsilon_0$.
Indeed, an increase in $A$ makes the tunneling barrier through
the Dirac gap $2\Delta_0$ shorter, and reduces the action 
under the barrier.
The other way of understading the apparent delocalization of the wave functions
for large $A$ is to notice that large potential 
amplitude results in strong oscillations of the potential term in 
Eq.~(\ref{H-nonint-gauged}). These oscillations effectively average this term to 
zero in the limit $A\to \infty$, in which case the minigaps become small
as $A^{-1/2}$. The latter follows from the large--$A$ asymptotic behavior of the 
Bessel functions for the weak coupling minigaps (\ref{minigaps-nonint})
that become valid in this limit.

\section{Bosonization}
\label{sec:boson}

The Hamiltonian (\ref{H}) 
is bosonized\cite{Stone,Kane'97,Egger'97,Talyanskii'01,Levitov'01} 
by virtue of the massive Thirring --- quantum sine-Gordon duality,
\cite{Coleman,Haldane} 
by representing the fermionic operators (\ref{psi-alpha}) 
as nonlocal combinations of bose fields
$\psi_{\alpha} \simeq (2\pi a)^{-1/2}\, e^{i\Theta_{\alpha}}$.
The conjugate momenta
\be \label{Pi-alpha}
\Pi_{\alpha} = {\hbar\over \pi v} \partial_t \Theta_{\alpha}
\ee
obey the canonical relations
$\lb \Pi_{\alpha}(x), \, \Theta_{\beta}(y) \rb_{-} 
= -i\hbar\delta_{\alpha\beta}\,\delta(x-y).$
The result is the Lagrangian 
\bea
\nonumber
\L & = & {\frac{\hbar v}{2\pi}} 
\sum_{\alpha=1}^4 \lp 
\frac1{v^2}\lp \partial_t \Theta_{\alpha} \rp^2 
- (\partial_x \Theta_{\alpha})^2  
- {2\Delta_0\over \hbar v a}  \cos2\Theta_{\alpha} \rp
\\ &-& 
\ds{\frac12}
\int \! dx' \, \rho(x) V(x-x')\rho(x')
- \rho \lf U(x) + \mu\rf \,,
\label{L-j}
\eea
where the density 
$\rho(x) = \sum_{\alpha=1}^4 \rho_{\alpha} 
= \sum_{\alpha=1}^4 \ts{\frac1{\pi}}\, \partial_x \Theta_{\alpha}$,
and $\mu$ is the chemical potential ($\mu=0$ at half-filling).
The Gaussian part of the action is diagonalized 
by the combination of the 
unitary transformation\cite{Talyanskii'01,Levitov'01}
\begin{eqnarray}
\label{transf}
\lp \matrix{\Theta_1\cr \Theta_2\cr \Theta_3\cr \Theta_4}\rp
=\frac12
\lp \matrix{1 & 1 & 1 & 1\cr 1 & -1 & 1 & -1\cr
1 & -1 & -1 & 1\cr 1 & 1 & -1 & -1 }\rp
\lp \matrix{\theta^0\cr \theta^1\cr \theta^2\cr \theta^3}\rp 
\end{eqnarray}
with the density (\ref{rho}) being a gradient of the charge mode $\theta^0$,
\be
\label{rho-bosonized}
\rho(x) = \textstyle{\frac{2}{\pi}}\, \partial_x \theta^0 \,,
\ee
and of the gauge transformation [cf. Eq.~(\ref{psi-gauge})]
\be \label{gauge}
\Theta_{\alpha}(x) \to 
\Theta_{\alpha}(x) - \textstyle\frac1{\hbar v}\int^x K^{-1} (U-\mu)\, dx' 
\ee
that leaves $\theta^a$ intact, and shifts 
\be \label{theta-gauge}
\theta^0 \to \theta^0 - \textstyle\frac2{\hbar v}\int^x K^{-1} 
(U-\mu)\, dx'  \,.
\ee
Here the charge stiffness 
\be \label{K}
K_q = 1 + 4\nu V(q) \,, \quad \nu = {1\over \pi \hbar v} \,.
\ee
Below we drop the (irrelevant) logarithmic dependence of the 
stiffness $K_q$ on the momentum, assuming a constant value
$K\!\equiv\! K_{q\sim 1/l_{\rm ch}}$, where $l_{\rm ch}\sim l_s$ is the 
charge soliton size (screening length for the Coulomb interaction).
Using $e^2/\hbar v\simeq 2.7$ we estimate $K\simeq 40$ for $l_{\rm ch} \sim 1\,\mu$m
for the stand-alone tube; $K\simeq 10$ for the tube 
placed on a substrate with a dielectric constant 
$\varepsilon=10$.   

The nonlinear part of the action (\ref{L-j}) is transformed 
using the identity 
$\sum_{\alpha=1}^4 \cos 2\Theta_{\alpha} = 4 \F(\theta^0, \theta^a)$, 
where 
\be
\label{f} 
\F \lp\theta^0, \,
\theta^a\rp
= \cos\theta^0 \cdot \prod_{a=1}^3 \cos\theta^a +  
\sin\theta^0 \cdot \prod_{a=1}^3 \sin\theta^a  , 
\ee
yielding the Lagrangian \cite{Talyanskii'01,Levitov'01}
\be \label{L'}
\L' = \L_0 + \L_{\rm bs} \,,
\ee
\bea 
\nonumber
\L_0   &=&  {\hbar v\over 2\pi} 
\lf
\frac1{v^2}(\partial_{t} \theta^0)^2 - K\, (\partial_{x}\theta^0)^2 
\rf
\\ 
& + & 
{\hbar v\over 2\pi} 
\sum_{a=1}^3 \lf \frac1{v^2}(\partial_t\theta^a)^2 - (\partial_x\theta^a)^2 \rf , 
\label{L0}
\\
\L_{\rm bs} & = & -  {\hbar v g_0\over \pi}  
\F(
\theta^0 + 2\tilde\mu\ksaw x - 2\tilde A \sin \ksaw x, \theta^a). \quad 
\label{LF}
\eea
The Lagrangian (\ref{L'})
describes one stiff charge mode $\theta^0$ and three
soft flavor modes $\theta^a$,  
nonlinearly coupled by 
\be \label{def-lambda0}
g_0={4\Delta_0\over \hbar v a} = {4 \over a^2}{\Delta_0 \over D} \,,
\ee
where $D$ is the bandwidth (\ref{D}), and the dimensionless quantities
\be \label{tilde-A}
\tilde A = {A\over K\epsilon_0} \,, 
\quad \tilde \mu = {\mu \over K\epsilon_0} 
\ee
are introduced in a way similar to that of Eq.~(\ref{bar-A}), 
with the energy scale $\epsilon_0$ defined in Eq.~(\ref{epsilon0}).
The difference between Eqs.~(\ref{tilde-A}) and (\ref{bar-A})
is in the {screening} (by a factor of $1/K$) of external fields 
$U(x)$ and $\mu$ by the interacting NT system.
When $g_0=0$, the electron number density (\ref{rho-alpha})
parametrized by $m$ varies continuously with $\tilde \mu$ as 
\be \label{mu-nu}
m = 2{\tilde \mu} \,. 
\ee
In this case the system is a compressible scale-invariant Tomonaga-Luttinger 
liquid of the four flavors regardless of the electron interaction strength.\cite{FSA}
The nonlinear term (\ref{LF}) breaks charge-flavor separation by binding
$\theta^a$ with $\theta^0$.
It can become relevant at commensurate densities, yielding incompressible states.

Note that the Lagrangians (\ref{L-j}) and (\ref{L'})
are SU(4) invariant, as is the original Hamiltonian $\H$.
This invariance is not explicit in the adapted notation.
It will manifest itself below on the level of renormalization.

\section{Phase soliton method}
\label{sec:int}

\subsection{Electron as a composite soliton}
\label{sec:switch}

\nin
Evaluation of excitation gaps requires the knowledge of how an added
electron is represented in the language of the coupled bosonic fields 
(\ref{transf}).
Here we first discuss this issue in a simpler case of a stand-alone tube,
when no potential is applied: $U(x)\equiv 0$.
We follow the framework of Levitov and Tsvelik,\cite{Levitov'01}
emphasizing the details that will be important in the rest of the work.

The idea is that when interactions are strong ($K\gg 1$),  
the action (\ref{L'}) is dominated by the charge sector,
which justifies optimization of (\ref{L'}) in a saddle point fashion,  
with $K^{1/2}$ playing the role of $\hbar^{-1}$.
In particular, the soft neutral modes $\theta^a$ 
adjust themselves to provide an effective potential 
for the charged mode $\theta^0$ in such a way that the cost of adding charge
is minimal. For the most part one can regard the charge mode as classical, since
its quantum fluctuations are suppressed by 
$K^{-1/2}$ (and can be further included as corrections to the scaling laws as 
described in Sec.~\ref{sec:fluct} below), 
whereas the neutral fluctuations strongly renormalize 
the coupling (\ref{def-lambda0}), with its renormalized value $g$
given by Eq.~(\ref{g-ren}) below.

As a result,\cite{Levitov'01}
an electron added to a half-filled nanotube is a {\it composite soliton}
characterized by the two length scales: the flavor $l_{\rm fl}$, and the charge 
$l_{\rm ch} \sim K^{1/2} l_{\rm fl}$, so that $\lch \gg \lfl$.
The neutral modes $\theta^a$ add a particular flavor 
to the electron, and, naturally, 
flavor is bound to charge by the non-linear term (\ref{LF}).
Adding an electron of a given polarization $\alpha$ amounts to 
changing $\Theta_{\alpha}$ by $\pi$.
According to the transformation (\ref{transf}), in the charge-flavor basis
this results in a configuration in which all 
the fields $\theta^0, \ \theta^a$ change by $\pm\pi/2$. 
However, given the separation of scales $l_{\rm ch} \gg l_{\rm fl}$, 
solitons of the flavor modes $\theta^a$  
appear as sharp steps on the scale of $l_{\rm ch}$ that  ``switch'' right in the 
middle of the charged soliton. 
The Coulomb interaction between electrons corresponds to overlap of 
charge soliton tails within the screening length $\lch$, 
whereas the fermionic exchange corresponds to overlap of the (shorter) 
flavor solitons.
In the low-density regime $\bar\rho \lfl \ll 1$, 
overlap of the charge-soliton tails 
maintans quasi-long-range order of a 1d Wigner crystal, whereas the 
effects of exchange are exponentially suppressed.

\begin{figure}[t]
\includegraphics[width=3.5in]{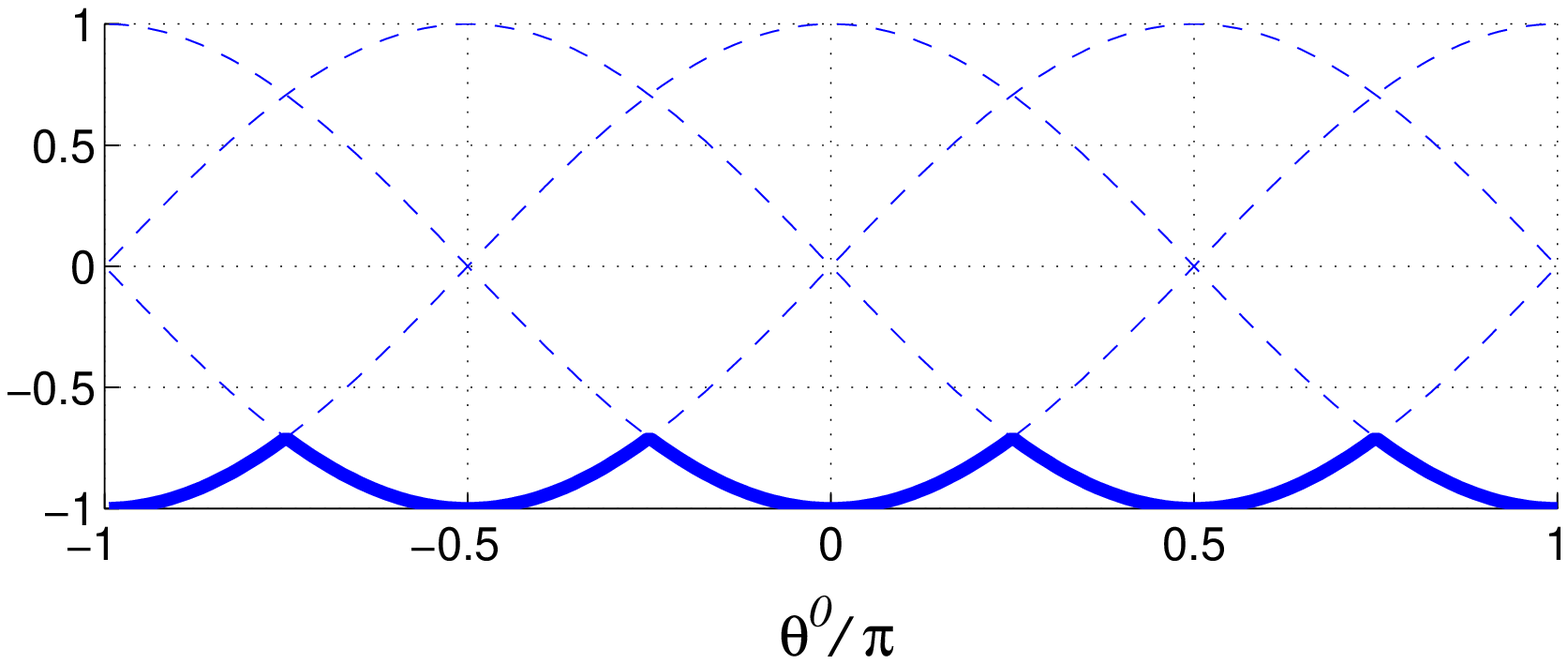}
\includegraphics[width=3.5in]{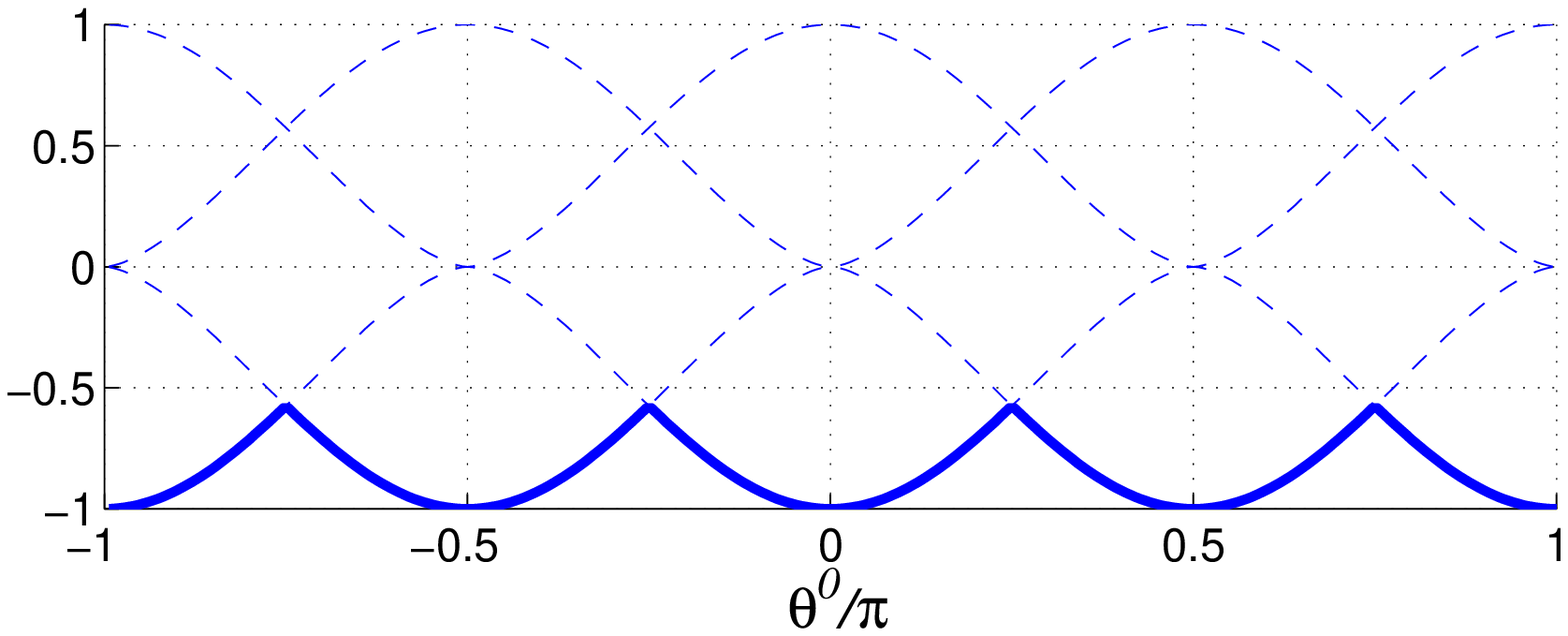}
\caption[]{(Color online)
Classical $\bar \F_{\rm cl}(\theta^0)$ (top) and renormalized 
$\bar \F(\theta^0)$ (bottom) 
effective potentials (shown in bold) for the charge mode
in the case of $K\gg 1$, Eqs.~(\ref{f-bar-cl}) and (\ref{f-bar})
}
\label{fig:f}
\end{figure}

Technically, the effect of the neutral modes is to 
provide Fourier harmonics quadrupling in the nonlinear term  (\ref{LF}),
$\F(\theta^0,\theta^a) \to \bar\F(\theta^0) \sim \cos 4\theta^0$. 
The optimization leading to the effective 
potential $\bar\F(\theta^0)$ is shown in Fig.~\ref{fig:f}.
On the classical level, minimizing $\F(\theta^0,\theta^a)$ over
the values $\theta^a = n_a \pi/2$ with integer $n_a$ yields
the potential
\be \label{f-bar-cl}
\bar \F_{\rm cl}(\theta^0) = \min_{\{ \theta^a = n_a \pi/2\} } 
\F (\theta^a , \ \theta^0) = \min \lf \pm \cos\theta^0 , \ \pm \sin\theta^0\rf 
\ee
for the charge mode (top panel of Fig.~\ref{fig:f}).
The switching between the branches of $\pm\cos\theta^0$ and $\pm \sin\theta^0$
occurs via the neutral fields changing by $\pm \pi/2$
right in the middle of the slow charge soliton, when
$\theta^0$ has changed by $\pi/4$.
One obtains the classical soliton of the charge mode  
of the form\cite{Levitov'01}
\be \label{sol-LT}
\theta_{\rm cl}^0(x)=\lf 
\matrix{
2\cos^{-1} \tanh(u - x/l_{\rm ch}^{(\rm cl)}) \,, & x<0 \,, \cr
\frac{\pi}{2} - 2\cos^{-1}\tanh(u+x/l_{\rm ch}^{(\rm cl)}) \,, & x>0 \,, 
}
\right.
\ee
where $l_{\rm ch}^{(\rm cl)} = (K/g)^{1/2}$ and $\tanh u = \cos \pi/8$.

Quantum fluctuations of the neutral sector renormalize the bare
potential $\sim g_0 \bar\F_{\rm cl}(\theta^0)$ 
by providing it with the scaling dimension
\be \label{gamma}
\gamma=8/5 \,.
\ee
As a result, the effective potential for the charge sector 
\be \label{f-bar}
\matrix{
\L_{\rm bs}\to -\lp{\hbar v \over \pi}\rp V_{\rm charge} \,, \quad
V_{\rm charge} = g \bar \F(\theta^0) \,, \cr
\bar \F(\theta^0) = - \lb - \bar\F_{\rm cl}(\theta^0)\rb^{\gamma} \,,
}
\ee
where the renormalized nonlinear coupling 
\be \label{g-ren}
g \simeq \frac1{a^2} \lp {\Delta_0 \over D}\rp^{\gamma} \,,
\ee
and the cutoff $D$ defined in Eq.~(\ref{D}).
The potential $\bar \F(\theta^0)$ is plotted in the lower panel 
of Fig.~\ref{fig:f}. 
The result (\ref{f-bar}) is justified 
in the adiabatic approximation $\lch\gg\lfl$ when the potential (\ref{f-bar-cl})
is slow on the scale $a < l < \lfl$ on which quantum fluctuations of the 
fields $\theta^a$ accumulate 
(for renormalization procedure see Sec.~\ref{sec:fluct} below).

Although the difference between 
the form of the potential (\ref{f-bar-cl}) 
[considered in Ref.~\onlinecite{Levitov'01}], 
and the renormalized one, $\bar\F(\theta^0)$ [Eq.~(\ref{f-bar})],  
is not crucial qualitatively,\cite{Novikov-unpub} 
it can be important for quantitative considerations.
Namely, it affects both the soliton energy, its size, 
$\lfl$,  and the effective screening length, $\lch$.
Unfortunately, the analytic form for the 
charge mode soliton with the potential term (\ref{f-bar}) is unavailable.
One way to find this soliton is to
approximate the potential (\ref{f-bar}) 
by a piecewise-quadratic function
[this procedure works well\cite{Levitov'01} for the soliton in the nonrenormalized
potential (\ref{f-bar-cl})]. 
This yields 
\be \label{sol-DN}
\theta^0(x) \approx \theta_{\rm quad}^0(x)=\lf 
\matrix{
\frac{\pi}{4} e^{x/\lch} \,, & x<0 \,, \cr
\frac{\pi}{2} - \frac{\pi}{4} e^{-x/\lch} \,, & x>0 \,, 
}
\right.
\ee 
with a charge soliton size 
\be \label{lch-DN}
\lch=\lp {K/\gamma g}\rp^{1/2} \,,
\ee
which is a factor $\approx 1.26$ shorter compared to that in Eq.~(\ref{sol-LT}).

Physically, strong Coulomb repulsion in the presence of the curvature $\Delta_0$
of electronic dispersion qualitatively changes the conventional
``spin-flavor separated'' Luttinger-liquid behavior governed by the 
Gaussian action (\ref{L0}).
At the Luttinger fixed point, all the modes enter on equal footing,
and the dispersion curvature is irrelevant at high density.
At low density, the saddle point of the total action (\ref{L'})
describes the crossover to the Wigner crystal, in which flavor is
bound to charge.
The neutral and the charge sectors 
play very different roles in the Wigner crystal regime: 
the former creates the effective potential for the latter.
This effective potential (\ref{f-bar}) has a period $\pi/2$ 
that is four times smaller than that of the potential (\ref{LF}) 
with {fixed} $\theta^a$; the lowest harmonic in the Fourier expansion 
\be \label{f-fourier}
\bar \F(\theta^0) = {\rm const} - \sum_{n=1}^{\infty} f_n \cos(4n\theta^0) 
\ee
is $\cos 4\theta^0$.
The purpose of this period reduction is to lower the Coulomb energy by
splitting the $\theta^0 \to \theta^0+2\pi$ 
excitation that carries charge $4e$ 
[according to Eq.~(\ref{rho-bosonized})] 
and is a flavor singlet, 
into four subsequent excitations each carrying a single fermion
of a unit charge and of a particular flavor.

Practically, to obtain a smooth approximate form of the charge soliton,
and to estimate its energy that dominates
the cost of adding an electron, 
here we keep the most 
relevant harmonic in the Fourier series (\ref{f-fourier}),
\be \label{F=cos4}
\bar\F(\theta^0) \approx -f_1 \cos 4\theta^0 \,, 
\quad f_1 \approx 0.178 \,.
\ee
The approximation (\ref{F=cos4}) overestimates the 
classical charge soliton energy by 10\%:
\be \label{Echsolapprox}
E_{\rm ch. sol.}^{\rm approx.} = c_1 {\hbar v\over \pi} \sqrt{Kg} \,, \quad 
c_1 = \sqrt{4f_1} \approx 0.844 \,, 
\ee
compared with the energy in the full potential (\ref{f-bar}):
\bea \label{Echsol}
E_{\rm ch. sol.} &=& c_\infty {\hbar v\over \pi} \sqrt{Kg} \,, \\ 
c_\infty &=& \sqrt{8}\int_0^{\pi/4}\! d\theta^0 \,\sqrt{1-\cos^\gamma\theta^0}
\approx 0.752 \,.
\nonumber
\eea
The charge soliton in the potential (\ref{F=cos4}) takes the form 
\be \label{sol-approx}
\theta^0_{\rm approx}(x) = \ts{\frac12} \cos^{-1}\tanh(-x/\lch')
\ee
with the rescaled size 
$\lch' = \frac14 (K/f_1 g)^{1/2}\approx l_{\rm ch}^{\rm (cl)}/1.69$.
Note that, due to strong renormalization by the neutral fluctuations,
the potential (\ref{f-bar}) is steeper than its classical counterpart
(\ref{f-bar-cl}).
This leads to {\it shorter} charge soliton length than that of a classical 
solution (\ref{sol-LT}), as both of the approximations, 
Eqs.~(\ref{sol-DN}) and (\ref{sol-approx}), indicate.

Remarkably, the difference between the Luttinger liquid 
[described by the Gaussian Lagrangian (\ref{L0})],
and the Wigner crystal
will become crucial after adding the external potential $U(x)$.
Depending on the relation between its period, $\lsaw$, and 
the ``size of an electron'', $\lfl$, either the Gaussian saddle point 
($g_0\to 0$) of the action (\ref{L'}),
or the Wigner-crystal saddle point described above will play out.
The latter one applies in the regime $\lsaw \gg \lfl$ 
considered below in Section~\ref{sec:strong}. There we show that 
the system effectively behaves as that of a single mode.

\subsection{Incompressible states of the classical bosonic fields}
\label{sec:sol4}

The above discussion of the transition to the Wigner crystal 
suggests that, already on the classical level, the nonlinear action 
(\ref{L-j}) or (\ref{L'}) captures important physics even on the 
short length scale (smaller than the inter-particle distance!). 
This intuition can be readily extended onto the $U\neq 0$ case.
By adding the external potential (\ref{SAW}), below we use the 
classical Hamiltonian 
\be 
\matrix{
\H_{\rm cl}[\Theta_{\alpha}] = 
\ds{1\over 2}\sum (\partial_{x}\Theta_{\alpha})^2 
+ \ds{{K-1\over 8}}\, \lp\sum \partial_{x}\Theta_{\alpha}\rp^2 
\cr 
+ \frac{1}{4}g_0 \sum \cos 2\Theta_{\alpha} 
+ \ds{\frac1{\hbar v}}\, \sum\partial_{x}\Theta_{\alpha}
\cdot \lp {U(x)}\!-\!\mu \vphantom{1^{2^3}}\rp ,
}
\label{H-bosonized-j}
\ee
that follows from the Lagrangian (\ref{L-j}),
to numerically illustrate how incompressible states appear.
For convenience here we work in the original basis $\Theta_{\alpha}$.

\begin{figure}[b]
\includegraphics[width=3.5in]{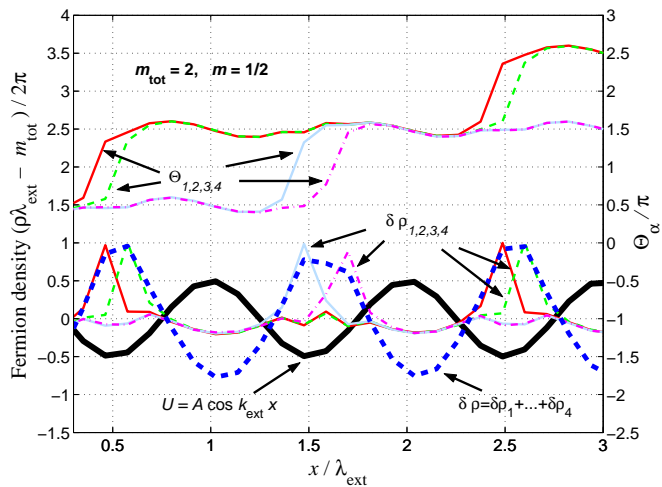}
\caption[]{(Color online)
Classical ground state of the Hamiltonian (\ref{H-bosonized-j}) 
for the density $m=1/2$.
Bold solid line marks the external potential.
Fine lines in the upper part denote the 
solitons of the fields $\Theta_{\alpha}$ found 
by the numerical minimization of Eq.~(\ref{H-bosonized-j})
for $\tilde\mu=1/4$, $K=10$, $g_0=12 \ksaw^2$, $A=K\epsilon_0$.
The corresponding fine lines in the lower part denote the flavor densities
$\delta\rho_{\alpha} = \rho_{\alpha} - m/\lsaw$ counted from their  
average value $m/\lsaw$. Bold dashed line is the total density
counted from its average value $4m/\lsaw$.
Note the period doubling for  $\Theta_{\alpha}(x)$.
}
\label{fig:sol4}
\end{figure}

Consider the simplest fractional density, $m=1/2$, corresponding to 
the chemical potential $\tilde\mu=1/4$ [cf. Eq.~(\ref{mu-nu})]. 
Its classical ground state is an incompressible configuration in which the solitons 
of the fields $\Theta_{\alpha}$ occupy every other potential minimum,
as shown in Fig.~\ref{fig:sol4}. This Figure is a result of the 
numerical minimization of the Hamiltonian (\ref{H-bosonized-j})
with respect to the fields $\Theta_{\alpha}$.
In agreement with the general theory\cite{Pokrovsky} (developed for a
single-mode system), for the fraction $m=p/q$ the 
density should have the period $q\lsaw$. In our case of the four modes,
naturally, the period of the density  
$\frac1\pi \partial_x\Theta_{\alpha}$ in each mode 
is $q\lsaw$,  with $q=2$ in Fig.~\ref{fig:sol4}.
Note that, since the {\it total} density $\mtot=2$ is integer, 
the charge density period coincides with that of $U(x)$. 
In the absence of $U(x)$, all the solitons of $\Theta_{\alpha}$ 
would be equally separated from each other due to the Coulomb repulsion.
The finite$-U$ configuration shown in Fig.~\ref{fig:sol4} 
is the result of an interplay between the mutually repelling solitons 
and a confining periodic potential.
We note that the fermionic exchange of the original problem (\ref{H})
is manifest here already on the classical level,
in the fact that solitons of the same flavor repel each other stronger, 
and, therefore, do not occupy the adjacent potential minima.

\subsection{Phase soliton method for a single mode}
\label{sec:phasesoliton}

To systematically study excitation gaps over commensurate configurations,
we employ and generalize the phase soliton method.
Representing an excitation in a commensurate phase by a
phase soliton was utilized in the past in various contexts. 
The model of locking a system into a commensurate state 
was first suggested in the work of Frenkel and Kontorova.\cite{Frenkel}
Later, it was re-discovered and solved by 
Frank and van der Merwe \cite{FvdM} in the context of atoms ordering
on crystal surfaces, and by Dzyaloshinskii \cite{Dzyal} 
describing a transition to the state with a helical magnetic structure.
The general theory of commensurate-incommensurate phase transitions 
has been finalized by Pokrovsky and Talapov.\cite{Pokrovsky}
The result of these investigations is a finite or infinite sequence of 
gaps (``devil's staircase'') corresponding to commensurate  states.
\cite{Bak}

The essense of the phase soliton method can be demonstrated on the
example of the single-mode system.
Consider the model Hamiltonian of the form 
\be \label{H1}
\H_1 = \frac{K}2 (\partial_x\theta)^2 
+ g \cos (\beta\theta + m \ksaw x + b \sin \ksaw x) 
\ee
[cf. Eqs.~(\ref{L-j}), (\ref{L'}), or (\ref{H-bosonized-j}),
albeit with only one mode present]. 
To make contact with the nanotube system, 
the parameters in the Hamiltonian (\ref{H1}) have a similar meaning
to those in 
the action (\ref{L-j}) transformed by (\ref{gauge}),
with $\beta=2$ and only $\Theta_1$ present. The Hamiltonian (\ref{H1}) 
is the continuous form of the Frenkel-Kontorova model\cite{Frenkel}
that describes locking of the compressible lattice of interacting 
particles (e.g. a chain of atoms) by the external potential of the period $\lsaw$
(e.g. the substrate potential for the atomic chain), as is schematically
shown in Fig.~\ref{fig:soliton-sketch}(a).
The phase field $\theta$ represents the deviations from 
equilibrium positions along the chain, 
with changing $\theta$ by $2\pi/\beta$ 
corresponding to a shift of the chain by its period.
The gradient $\beta \partial_x \theta/2\pi$ thus has a meaning of the excess 
particle density near point $x$.
The coupling between the particles and the potential is represented by $g$,
whereas the lowest harmonic of the potential is $\propto b\cos\ksaw x $. 
The crucial parameter is the 
period ratio $m$ (number of particles per period $\lsaw$). 

External potential can lock the chain into incompressible states
when $m$ is either an integer, or a simple fraction $m=p/q$.
An excitation (an incommensuration) over such a commensurate state 
corresponds to adding an extra particle
to the system. This changes the positions of other particles in the chain
due to their mutual interaction.
When the coupling $g$ is small, such a change occurs over many periods $\lsaw$
[as illustrated in Fig.~\ref{fig:soliton-sketch}(c)].

In the language of the continuous model (\ref{H1}), 
the incommensuration is represented by the soliton-like change of $\theta$
by $2\pi/\beta$, with the excitation gap given by the soliton energy. 
The effective theory for the excitation follows from
the Euler-Lagrange equation for (\ref{H1}), $\delta \H_1 /\delta\theta = 0$,
practically found by expanding in the powers of $g$,
\be \label{PT-expansion}
\theta = \bar \theta + {\theta}^{(1)} + ... + {\theta}^{(n)} + ... \,,
\quad \theta^{(n)} = {\cal O}(g^n) \,.
\ee
Here $\bar\theta$ is constant in the commensurate phase,
acquiring a slow coordinate dependence $\bar\theta(x)$ in an excitation
({\it phase soliton}).
When $m$ is integer, the zeroth order term $\theta=\bar\theta$ in the expansion
(\ref{PT-expansion}) is enough:
Averaging over a period $\lsaw$, the effective Hamiltonian 
is $K(\partial_x \bar\theta)^2/2 + g J_m(b) \cos \beta\bar\theta$, 
and the excitation is described by 
the sine-Gordon soliton of a slow mode $\bar\theta(x)$ that 
interpolates by $2\pi/\beta$ over a few $\lsaw$.

For the fractional density $m=p/q$, averaging 
$g\cos (\beta\bar\theta + m \ksaw x + b\sin \ksaw x)$ over $q\lsaw$ gives zero
for constant $\bar\theta$. This indicates that the effective potential
for $\bar\theta$ is of the higher order in the coupling $g$. 
Using the expansion (\ref{PT-expansion}), one obtains
the lowest order Hamiltonian for the phase soliton $\bar\theta(x)$ of the form 
$K(\partial_x\bar\theta)^2/2 + g_q(b)\cos \beta q\bar\theta $, $g_q\sim g^q$.
The corresponding soliton energy estimates the excitation gap. 
One also has to average over the thermal or quantum fluctuations 
to assess the relevance of the potential term $\sim \cos \beta q\bar\theta$,
in which case the (omitted) 
overall coefficient in front of the Hamiltonian (\ref{H1})
becomes important, as it controls the size of fluctuations.

\subsection{Phase soliton method for a nanotube}
\label{sec:weak-strong}

Below we show that the case of multiple modes 
is qualitatively different from a conventional 
single mode situation described above. 
This difference manifests itself in the presence 
of the two regimes of coupling of the NT electron system to the external potential.
Physically, the choice of the regime depends on 
whether the electron wavefunctions overlap with each other substantially,
 --- that is, whether the physics is determined 
by the quantum interference with fermionic exchange being important, 
or by the Coulomb repulsion between nearly point-like electrons, 
in which case the exchange effects are negligible.

Technically, the difference between the regime of Bragg diffraction 
in a Luttinger liquid (``weak coupling''), or of pinning of the Wigner crystal
(``strong coupling'') is determined by the saddle point of the 
nonlinear action (\ref{L'}). In particular,
if the neutral modes switch fast on the scale on which the 
argument of the potential (\ref{LF}) changes appreciably,
the Wigner-crystal saddle point (Sec.~\ref{sec:switch}) is selected,
and the system is described in terms of the single (charge) mode,
with the problem reduced to that described in Sec.~\ref{sec:phasesoliton}
above. If, on the other hand, the neutral modes are slow, 
the weak-coupling saddle point 
(expansion around $g_0=0$) is the relevant one.

To define the regimes of coupling we need to compare the size  
$\bar\lfl$ of the phase soliton $\bar\theta^a(x)$ of the flavor sector 
(size of the added electron) with the potential period $\lsaw$. 
For now, we assume that $\bar\lfl$ is known.
The scale $\bar\lfl$ will be later 
obtained self-consistently as a result of integrating 
over the quantum {fluctuations} (performed for different cases in 
Sections~\ref{sec:fluct}, \ref{sec:weak} and \ref{sec:strong}).

\subsubsection*{Weak coupling} 

In the weak coupling regime the flavor soliton tails, 
which correspond to fermions of the same flavor, overlap.
Physically, it means that the system ``knows'' that it is composed 
of the particles of different flavors since 
the role of exchange is important. An extreme example of this regime 
is a non-interacting system 
in which fermions of the same flavor 
effectively repel due to the Pauli principle, whereas 
fermions of different flavors do not notice each other.

Strong overlap between the soliton tails in a commensurate 
configuration requires large $\lfl \gg \lsaw$, or, equivalently,
small coupling $g_0$ in Eq.~(\ref{L'}). 
The system is close to the Luttinger liquid, i.e. it is compressible for
almost all densities apart from a few commensurate ones, for which small minigaps
open up. This regime has been 
considered in Ref.~\onlinecite{Talyanskii'01}
for integer $m$.
The smallness of $g_0$ warrants finding the effective Lagrangian 
\be \label{Lm-general}
\L_{m}[\bar \theta^0, \bar \theta^a] = \L_0[\bar \theta^0, \bar \theta^a]
+ \L_m^{\rm int}[\bar \theta^0, \bar \theta^a] 
\ee
for the phase modes $\bar\theta^0$ and $\bar\theta^a$ 
perturbatively in $g_0$ (similarly to the single-mode model case 
considered in Sec.~\ref{sec:phasesoliton} above), 
treating all the modes on equal footing. 
Namely, from the Lagrangian (\ref{L'}) with 
the chemical potential (\ref{mu-nu}) corresponding to the commensurate 
density (\ref{rho-alpha}) parametrized by $m=p/q$,
one finds the effective potential 
$\L_m^{\rm int}[\bar \theta^0, \bar \theta^a]$
for the phase modes to the lowest order in $g_0$ by
solving the time-independent Euler-Lagrange equations 
$\delta \L'/\delta \theta^0=0$ and  $\delta \L'/\delta \theta^a=0$,  
via the expansion 
\bea \label{bar-theta}
\matrix{
\theta^0 = \bar \theta^0 + {\theta^0}^{(1)} + ... + {\theta^0}^{(n)} + ... \,, 
\cr 
\theta^a = \bar \theta^a + {\theta^a}^{(1)} + ... + {\theta^a}^{(n)} + ... \,, \cr
{\theta^{0}}^{(n)} = {\cal O}(g_0^n)  \,, \quad 
{\theta^{a}}^{(n)} = {\cal O}(g_0^n)\,.
}
\eea
The first term in (\ref{Lm-general})
is the Gaussian Lagrangian (\ref{L0}) as a function of the slow phase modes,
and the obtained potential energy $\L_m^{\rm int}$ is of the 
order $g_0^q$ (before integrating over quantum fluctuations).

From the Lagrangian $\L_{m}[\bar \theta^0, \bar \theta^a]$
one finds the commensurate ground state in which the 
phase modes $\bar\theta^0$ and $\bar\theta^a$ are constant. 
Similarly to the electron added to a stand-alone tube being represented
by a composite soliton (Sec.~\ref{sec:switch}), 
an excitation over such a ground state is a 
{\it composite phase soliton} of 
the slow phase modes $\bar \theta^0(x), \ \bar\theta^a(x)$.
It can be found as a saddle point of 
the Lagrangian $\L_{m}[\bar \theta^0, \bar \theta^a]$ in a way similar to 
that described above in Sec.~\ref{sec:switch}.

The weak coupling limit is illustrated below by considering 
the simplest situation of {\it integer} density $m$, in which case the  
decomposition (\ref{bar-theta}) is trivial since it contains just one term 
for each mode. 
Hence $\L_{m}[\bar \theta^0, \bar \theta^a]$ is precisely the 
Lagrangian (\ref{L'}) written as a function of 
$\theta^{0,a} = \bar\theta^{0,a}$ with the chemical potential $\tilde\mu$
given by Eq.~(\ref{mu-nu}).

Technically, the weak coupling regime is characterized by 
the slow ``switching'' of the flavor modes $\theta^a$ 
on the scale on which the phase of the charge part of the potential energy 
$\L_m^{\rm int}[\bar\theta^0,\,\bar\theta^a]$ in the Lagrangian (\ref{Lm-general})
changes by $2\pi$. For integer $m$, 
the potential energy is just the nonlinear term (\ref{LF}) written as
a function of the slow modes $\bar \theta^0, \, \bar\theta^a$.
Thus we demand
\be \label{weak}
\matrix{
m\lfl \gg \lsaw  \,, \cr  
2\tilde A \sin \ksaw \lfl \gg 2\pi \,.
}
\ee
The first condition in Eq.~(\ref{weak})
requires the flavor excitation to be extended 
on the scale of the separation 
$4\bar\rho^{-1} = \lsaw/m$ between same-flavor fermions.
The meaning of the second condition will be made more clear below.
The weak coupling limit (\ref{weak}) 
results in the following approximation for the energy (\ref{LF}): 
\be \label{F-weak}
\matrix{
\F (\bar \theta^0 + m\ksaw x - 2\tilde A \sin \ksaw x ,\ \bar \theta^a)  \cr
\approx J_m(2\tilde A) \F \lp \bar \theta^0,\ \bar \theta^a\rp 
\to  J_m(2\tilde A) \bar\F_{\rm cl}(\bar \theta^0)  \,.
}
\ee
Here we discarded spatially oscillating terms 
(in other words, we averaged $\F$ over the period $\lsaw$),
and denoted by the arrow the soft mode ``switching'' 
that produced the optimized potential (\ref{f-bar-cl}).

The procedure (\ref{F-weak}) as written is allowed for the integer density $m$.
In the case when the density $m=p/q$ is a simple fraction 
one needs to utilize the phase soliton
approach to find the effective potential of order $g_0^q$ and then 
perform analogous optimization.
The weak coupling limit is considered in Sec.~\ref{sec:weak-integer}
for integer $m$ and in Sec.~\ref{sec:weak-fractional} for the simplest 
fractional density $m=1/2$.

\subsubsection*{Strong coupling}

In the opposite, strong coupling limit, the Coulomb interaction 
wins over the effects of the fermionic exchange.
In this regime the system ``forgets'' its four-flavor nature,
and effectively behaves as that of spinless Dirac fermions
with the total density (\ref{rho-alpha}).
Technically, 
the soft mode ``switching'' (denoted by the arrow below) occurs {\it before}
averaging over the period of the potential,
\bea
\nonumber
\F \lp \bar \theta^0 + m\ksaw x - 2\tilde A \sin \ksaw x ,\ \bar \theta^a\rp \\
\nonumber
\to \bar \F \lp \bar \theta^0 + m\ksaw x - 2\tilde A \sin \ksaw x\rp \\
\nonumber
= f_1\cos\lp 4\bar\theta^0 + \mtot\ksaw x -8\tilde A \sin \ksaw x\rp \\
\nonumber
+ f_2\cos\lp 8\bar\theta^0 + 2\mtot\ksaw x -16\tilde A \sin \ksaw x\rp + ... \\
\approx f_1 \cos \lp 4\bar\theta^0 + \mtot\ksaw x -8\tilde A \sin \ksaw x\rp \,.
\label{F-strong}
\eea
Here the Fourier coefficients $f_n$ are defined in Eq.~(\ref{f-fourier}),
and in the last line we used the approximation (\ref{F=cos4}).
Note the dependence of the resulting potential energy on the {\it total density} 
$\mtot=4m$. The condition for the saddle point (\ref{F-strong})
is
\be \label{strong}
\matrix{
m\lfl < \lsaw  \,, \cr  
2\tilde A \sin \ksaw \lfl < 2\pi \,.
}
\ee
This regime is considered below in Section~\ref{sec:strong}.

Now let us clarify the meaning of the second condition in 
Eqs.~(\ref{weak}) and (\ref{strong}).
Increasing the potential amplitude $A$ much beyond $K\hbar v/\lfl$
effectively averages out the nonlinear term and thus
draws the system into the weak coupling limit (\ref{weak}).
This situation is similar to the noninteracting case considered in 
Section~\ref{sec:nonint} above, where the approximation 
(\ref{minigaps-nonint}) becomes valid at large $A$.

To summarize, above we have generalized the phase soliton method 
to the multiple-mode case. After the appropriate 
effective action for the slow modes is chosen 
[either Eq.~(\ref{Lm-general}), or the one 
based on the potential energy (\ref{F-strong})], excitations are described  
by the corresponding phase solitons.
Due to the electron-hole symmetry of the original problem,
adding an electron to a commensurate state (phase soliton)
costs the same energy $\Delta_m$ as removing an electron from the same state 
(anti-soliton), while the sum of the two energies, $2\Delta_m$,
is the corresponding excitation gap.

\subsection{Effect of quantum fluctuations}
\label{sec:fluct}

\nin
To illustrate integration over quantum fluctuations,
below we renormalize the gap $\Delta_0$ for the stand-alone tube.
(Generalizations for the $U\neq 0$ will be made in subsequent Sections.)
Fluctuations of the neutral sector around the saddle point 
described in Sec.~\ref{sec:switch} above 
are governed by the Lagrangian (\ref{L'})
with fixed value of $\theta^0$.
Adiabaticity of the charged mode at $K\gg1$ ensures that 
$\theta^0 \simeq {\rm const}$
on the length scales $a < l < \lfl$ where quantum fluctuations of the 
flavor sector accumulate. 
With each neutral field 
contributing to the scaling dimension by 1/4, 
flavor fluctuations result in the scaling dimension
\be \label{gamma0}
\gamma_0 = 3/4
\ee
for the adiabatic charge mode potential (\ref{f-bar-cl}),
yielding  
\be \label{rg0}
V_{\rm charge}(l) =  V_{\rm charge}(a) \lp {l\over a}\rp^{-\gamma_0} ,
\quad  V_{\rm charge}(a) = g_0 \bar\F_{\rm cl}(\theta^0) \,,
\ee
where $l$ is the renormalization-group (RG) scale, and $a$ is the tube radius.
Since $\gamma_0<2$, the potential $V_{\rm charge}$ 
is relevant and grows.
The flow (\ref{rg0}) stops on the scale $l\simeq \lfl$
on which the potential energy becomes comparable to the kinetic energy of the flavor
sector, since the perturbative RG yields the law (\ref{rg0}) 
only while the renormalized coupling 
stays small.\cite{CallanSymanzik}
For the larger scales $l> \lfl$ the potential energy dominates and the problem
becomes classical.

The scale $\lfl$ has a twofold meaning.
First, it is the {\it correlation length} for the flavor,
estimated self-consistently from the balance of kinetic and potential terms 
\be\label{l0-self-consist}
V_{\rm charge}(\lfl) \simeq {1\over \lfl^2} \,.
\ee
Beyond $x\simeq \lfl$ the correlation functions of the $\theta^a$ fields decay 
exponentially, rather than in a power-law fashion.
From Eqs.~(\ref{rg0}) and (\ref{l0-self-consist}) we obtain
the renormalized potential (\ref{f-bar}) for the charge mode
with the coupling (\ref{g-ren}) and the scaling exponent (\ref{gamma}),
\be \label{gamma-gamma0}
\gamma = {2\over 2-\gamma_0} \,.
\ee
The second meaning of the scale $\lfl$ is the renormalized
{\it size of the flavor soliton} of the model (\ref{L'}) with fixed $\theta^0$.

Since the charge sector is stiff, 
the excitation gap $\Delta$ is dominated by the charge soliton 
energy. The latter can be now estimated 
classically [cf. Eq.~(\ref{Echsolapprox})] from the effective charge mode Lagrangian
\bea \nonumber
\L_{\rm charge} = 
\ds{\hbar v\over \pi} 
\lp 
\frac1{2v^2}(\partial_t\theta^0)^2 - \frac{K}2 (\partial_x\theta^0)^2
- V_{\rm charge} 
\rp  \\
\approx 
\ds{\hbar v\over \pi} 
\lp 
\frac1{2v^2}(\partial_t\theta^0)^2 - \frac{K}2 (\partial_x\theta^0)^2
- g f_1 \cos 4\theta^0 
\rp , 
\label{Lcharge}
\eea
where we used the saddle point approximation (\ref{f-bar}), (\ref{g-ren})
and (\ref{F=cos4}) outlined in Sec.~\ref{sec:switch} above.
In the limit $K\to\infty$ the effective Lagrangian (\ref{Lcharge}) 
yields the gap\cite{Talyanskii'01,Levitov'01}
\be \label{Delta-Kinf}
\Delta_{K\to\infty} \simeq K^{1/2} D^{1/5} \Delta_0^{4/5} \,.
\ee

At finite $K$ the problem of finding exact scaling laws in the nonlinear theory
(\ref{L'}) is far from trivial. Here we attempt to estimate the 
${\cal O}(\eta)$ corrections, 
\be \label{eta}
\eta = K^{-1/2} \,, 
\ee
to the $K=\infty$ scaling laws in Eq.~(\ref{Delta-Kinf}).
For that we note that in one-loop RG, 
one averages the nonlinear term $\L_{\rm bs}$ over 
the {\it independent} Gaussian fluctuations described by the 
Gaussian action $\L_0$ diagonal in charge and flavor. 
Due to the factorization (\ref{f}), the result of such an averaging 
can be represented in the form of the product
of the contributions of independent Gaussian theories,
with the terms of the action (\ref{L'}) scaling as
\be \label{action}
{K\over \lch^2} + {1\over \lfl^2} 
+ g_0 \lp{\lfl\over a}\rp^{-\gamma_0} \lp{\lch\over a}\rp^{-\eta/4} .
\ee
Minimizing the composite-soliton action of the form (\ref{action})
over {\it both} $\lch$ and $\lfl$ amounts to including the coupling 
between the flavor and charge fluctuations by the nonlinear term.
The variational estimate yields
\be
{\lfl \over a} \simeq K^{\zeta} \lp{D\over \Delta_0}\rp^{4\over 5-\eta} ,
\quad \lch \simeq K^{\frac12} \lfl \,, 
\quad  \ts{\zeta={\eta \over 10-2\eta}} 
\ee
for the correlation lengths, and produces the renormalized gap 
$\Delta \simeq \hbar v K/\lch$ that scales as
\be \label{Delta-K}
\Delta \simeq K^{\frac12-\zeta} D^{\frac{1-\eta}{5-\eta}}
\Delta_0^{\frac{4}{5-\eta}} \,.
\ee
The latter expression smoothly crosses over to the noninteracting case $K=1$,
$\Delta=\Delta_0$, and it also has a correct  
scaling (\ref{Delta-Kinf})  
in the classical limit $K\to \infty$.\cite{compare-LT}

\section{Weak coupling limit}
\label{sec:weak}

\subsection{Integer density $m=\pm 1, \pm 2, ...$}
\label{sec:weak-integer}

The saddle point (\ref{F-weak}) maps the problem of finding excitation gaps
onto that with $U\equiv 0$ and 
\be \label{def-lambda-A}
g_0 \to g_m \simeq g_0 J_m(2\tilde A)  \,.
\ee
Based on the results of Sections \ref{sec:switch} and \ref{sec:fluct}, 
one obtains the renormalized minigaps $2\Delta_m$, where
\be \label{minigaps-ren}
\Delta_m 
\simeq  K^{1/2-\zeta} D^{{1-\eta\over 5-\eta}} 
\lp \Delta_m^{(0)}(\tilde A) \rp^{4\over 5-\eta}  
\ee
[cf. Eq.~(\ref{Delta-K})], and where the bare minigaps 
\be \label{minigaps}
\Delta_m^{(0)}(\tilde A) = \Delta_0 \, |J_m(2\tilde A)| 
\ee
given by their noninteracting values (\ref{minigaps-nonint}) with 
the screened potential amplitude $\tilde A$ defined in Eq.~(\ref{tilde-A}).
Eq.~(\ref{minigaps-ren}) in the $K\to\infty$ limit corresponds to 
minigaps found in Ref.~\onlinecite{Talyanskii'01}.
The qualitative features of the noninteracting 
minigaps (\ref{minigaps-nonint}) persist in the interacting case.\cite{Talyanskii'01}
Namely, as a function of the screened potential
amplitude (\ref{tilde-A}), the minigaps (\ref{minigaps-ren}) oscillate,
vanishing at particular values of $\tilde A$.
However, minigaps (\ref{minigaps-ren}) are strongly enhanced 
in magnitude compared to  (\ref{minigaps-nonint}) due to electron interactions. 
Also the dependence of the minigaps on both the bare backscattering 
$\Delta_0$ and on the periodic potential amplitude $A$ 
has a characteristic power law behavior which, 
in the limit of strong interactions $K\gg 1$, is given by a universal
power law $4/5$ characteristic of the number of NT fermion polarizations
at half-filling.

What is the cost of adding electron's {\it flavor} on top of its charge?
According to the mapping (\ref{def-lambda-A}),
the problem is formally equivalent to that with $U\equiv 0$, so
the results of Ref.~\onlinecite{Levitov'01} apply:
On the energy scale below that 
of the frozen charge sector, the flavor sector
is governed by the effective SU(4)$\simeq$O(6)
Gross-Neveu Lagrangian\cite{Levitov'01,GN}
\bea \nonumber
\L_{\rm GN} &=& {\hbar v \over \pi} \int \! dx 
\lf \frac12 (\partial_{\mu}\theta^a)^2  - g_{\rm GN}
\sum_{a>b} \cos2\theta^a\cos2\theta^b \rf \\
&=& \hbar v \int \! dx \lf i\bar\chi_j \gamma_{\mu}\partial_{\mu}\chi_j - 
g_{\rm GN} (\bar \chi_j \chi_j)(\bar \chi_{j'} \chi_{j'}) \rf ,
\label{L-GN}
\eea
where $\chi_j$ are the Majorana fermions and 
the Gross-Neveu coupling in our case is $g_{\rm GN} \propto g_m$.
The excitations of the model (\ref{L-GN}) 
are {\it massive} relativistic particles transforming according to 
different representations of the O(6) group,\cite{Z}
with the mass scale physically originating due to effects of exchange.

\subsection{Fractional density $m=\frac12 $}
\label{sec:weak-fractional}

\nin
Using the phase soliton method outlined
in Sec.~\ref{sec:phasesoliton} above, 
we derive the effective phase mode Lagrangian 
\be \label{L012}
\L_{1/2}[\bar \theta^0, \bar \theta^a] = \L_0[\bar \theta^0, \bar \theta^a] 
+ \L_{1/2}^{\rm int}[\bar \theta^0, \bar \theta^a] 
\ee
given by the sum of the Gaussian part $\L_0$ [Eq.~(\ref{L0})] and  
the potential energy [see Appendix~\ref{app:12} for details]
\bea 
\nonumber
- \L_{1/2}^{\rm int} &=& 
{\hbar v g_0' \over 4\pi} 
\lf \vphantom{\sum_a^3}
(4-\kappa)v_{1/2}(2\tilde A) 
\F (2\bar\theta^0, \, 2\bar\theta^a) 
 \right. \\ && \left.
\nonumber
-\kappa  v_{1/2}(2\tilde A) \sum_{a} \cos 2\bar\theta^a \cos 2\bar\theta^0
 \right. \\ && \left.
+\kappa  u_{1/2}(2\tilde A) \sum_{a>b} \cos 2\bar\theta^a \cos 2\bar\theta^b
\rf . \quad 
\label{bar-L-12}
\eea
Here the function $\F$ is defined in Eq.~(\ref{f}) and the other quantities 
are defined in Appendix~\ref{app:12}.

Let us discuss the potential energy (\ref{bar-L-12}).
Its first term has scaling dimension 3 and is irrelevant. 
The second term of the potential (\ref{bar-L-12})
is responsible for the charge excitation gap $\Delta_{\rm ch, 1/2}$.
In the limit $K\gg 1$, this term has scaling dimension 
\be \label{gamma12}
\gamma_{1/2} = 1
\ee
and grows under the renormalization group flow. 
Finally, the third term is marginal. It describes the SU(4) {\it flavor} 
physics on the energy scale $\Delta_{\rm fl} \sim K^{-1/2} \Delta_{\rm ch,1/2}$.

The charge gap is found via
the saddle-point optimization of the second term of the potential (\ref{bar-L-12})
by the neutral sector in a way similar to that of 
Sec.~\ref{sec:switch} above. 
Adding an extra electron to the system (\ref{L012}) corresponds to 
a composite phase soliton in which the $\bar\theta^a$ fields
``switch'' by $\pi/2$ in the middle of a slow charged phase soliton,
at the point when $\cos2\bar\theta^0=0$.
(It is possible to show\cite{Novikov-unpub} that although at this very point 
the effective sine-Gordon coupling $\propto \cos2\bar\theta^0$ 
for the neutral sector vanishes, 
the flavor soliton has a finite size and energy. 
This is the case since away from the center of the
charged soliton, $\cos2\bar\theta^0 \neq 0$ giving the finite flavor scale.)

\begin{figure}[t]
\includegraphics[width=3.5in]{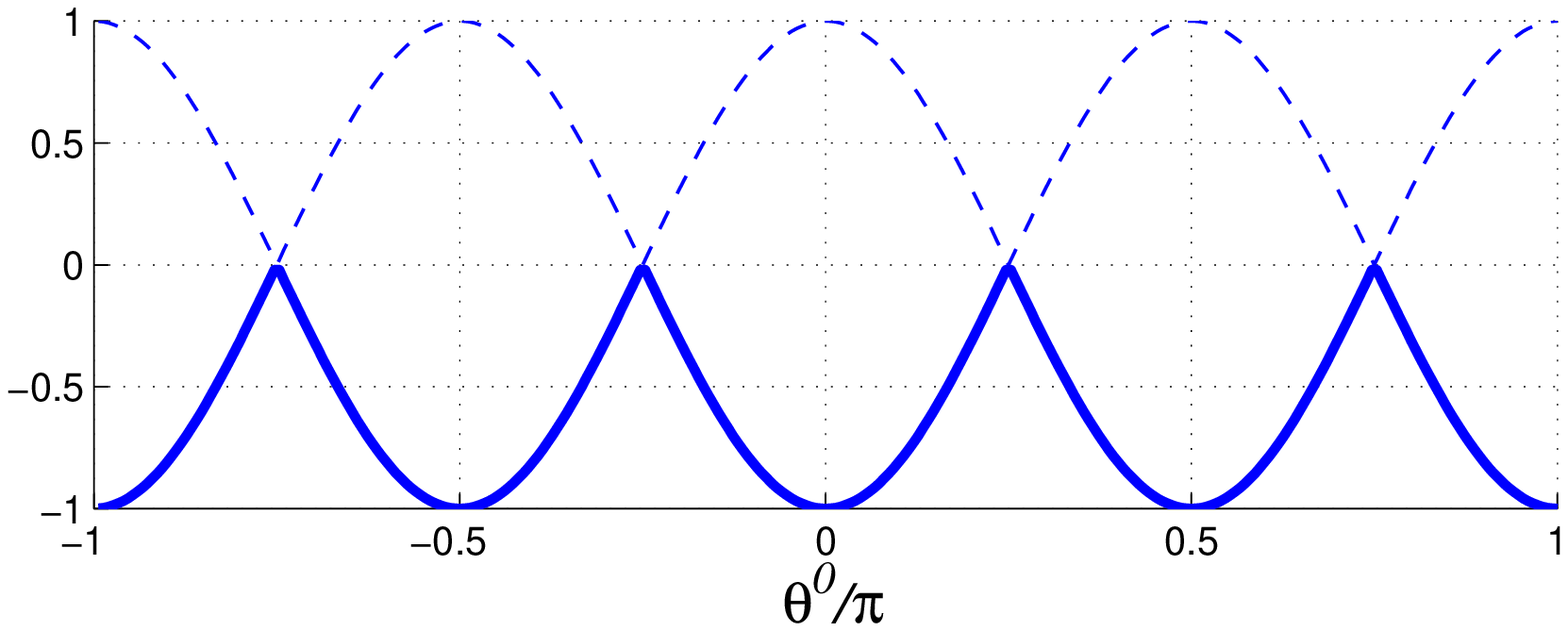}
\includegraphics[width=3.5in]{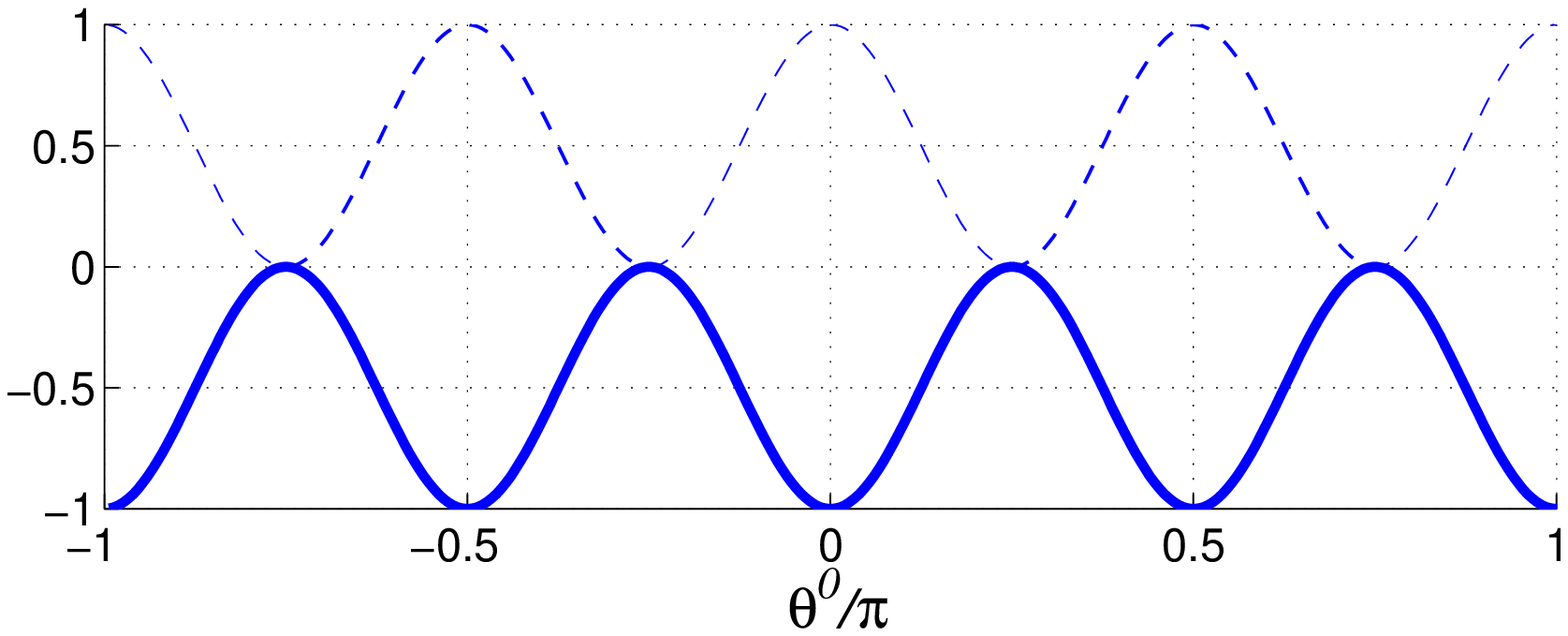}
\caption[]{(Color online)
Classical $\bar \F_{\rm cl}^{(1/2)}(\theta^0)$ (top) and 
renormalized $\bar \F^{(1/2)}(\theta^0)$ (bottom) 
effective potentials (shown in bold) 
for the charge mode at density $m=1/2$ 
in the case of $K\!\gg\!1$, Eqs.~(\ref{f12}) and (\ref{Vcharge-12})
correspondingly
}
\label{fig:f12}
\end{figure}

On the classical level, the neutral sector produces
the charge potential $\bar \F_{\rm cl}^{(1/2)}$ 
[top panel of Fig.~\ref{fig:f12}],
\be 
\sum_a \cos2\bar\theta^0 \cos2\bar\theta^a 
\to \bar \F_{\rm cl}^{(1/2)}(\bar\theta^0) = - \left| \cos2\bar\theta^0 \right| \,.
\label{f12}
\ee
By integrating over the flavor fluctuations 
we obtain the flow of the form (\ref{rg0}) with the scaling exponent (\ref{gamma12})
for the potential energy 
$g_{1/2} \bar \F_{\rm cl}^{(1/2)}(\bar\theta^0)$,
where
\be \label{g12}
g_{1/2} =  \ts{\frac14} \kappa g_0'  
|v_{1/2}(2\tilde A)|  \,.
\ee
From the self-consistency of the form (\ref{l0-self-consist}) 
we find the flavor scale
\be \label{l12}
\lfl_{,1/2} \sim {a\over \kappa |v_{1/2}(2\tilde A)|}  
\lp {\epsilon_0 \over \Delta_0}\rp^2 ,
\ee
[$a$ is the tube radius], the renormalized coupling
\be \label{g*12}
g_{1/2}^* \simeq {1\over \lfl_{,1/2}^2}  \propto g_{1/2}^2 \,,
\ee
and the scaling exponent
\be \label{gamma-gamma-12}
\tilde\gamma_{1/2} = 2 
\ee 
for the classical optimized potential (\ref{f12}).
As a result, the 
adjustment of the neutral sector yields the following effective potential
for the charge mode:
\be \label{Vcharge-12}
\matrix{
V_{\rm charge}^{(1/2)}   =  
g^*_{1/2} \bar\F^{(1/2)}(\bar \theta^0) ,  
\cr
\bar\F^{(1/2)} = -  \left| \bar \F_{\rm cl}^{(1/2)} \right|^{\tilde\gamma_{1/2}} 
= -\ts{\frac12} \cos4\bar\theta^0 + {\rm const.}   
}
\ee
Note that, due to the specific value of the scaling (\ref{gamma12})
[remarkably coinciding with the gap scaling for 
a single mode of  noninteracting fermions!],
the renormalized potential $\bar \F^{(1/2)}$ has only one Fourier harmonic,
hence the coefficient $1/2$  
in Eq.~(\ref{Vcharge-12})
is not approximate [as one could anticipate
from the analogous procedure leading to Eq.~(\ref{F=cos4})], but {\it exact}.
The functions $\bar \F_{\rm cl}^{(1/2)}$ and $\bar \F^{(1/2)}$ 
are shown in Fig.~\ref{fig:f12}.
They both have a period $\pi/2$ that corresponds to adding unit charge
according to Eq.~(\ref{rho-bosonized}). 
As a result, the charge soliton for $m=1/2$ in the limit $K\gg 1$  has the 
{\it exact} form (\ref{sol-approx}) 
with the size $\lch_{,1/2}|_{K\gg 1}=\frac14 (2K/g^*_{1/2})^{1/2}
\sim K^{1/2}/g_{1/2}$.

For large but finite $K$, the soliton sizes $\lch_{,1/2}$, $\lfl_{,1/2}$, and 
the renormalized minigap $\Delta_{\rm ch, 1/2}\simeq K \hbar v /\lch_{,1/2}$
can be found from the variational estimate of the form (\ref{action}),
\be
{K\over \lch^2_{,1/2}} + {1\over \lfl^2_{,1/2}} 
+ g_{1/2} \lp {\lfl_{,1/2}\over a}\rp^{-1} \lp {\lch_{,1/2}\over a}\rp^{-\eta} , 
\ee
yielding [$\eta\equiv K^{-1/2}$, $\kappa=(K-1)/K$]
\be \label{Delta-12}
\Delta_{\rm ch, 1/2} \simeq
K^{1-2\eta\over 2-2\eta} \, D\, 
\lb \kappa |v_{1/2}(2\tilde A)| 
\lp {\Delta_0 \over \epsilon_0}\rp^2\rb^{1\over 1-\eta} .
\ee
The value of the gap (\ref{Delta-12}) 
is strongly {\it enhanced} by the bandwidth $D$ 
[even stronger than for the integer$-m$ case of Sec.~\ref{sec:weak-integer}]
due to flavor fluctuations.
We also note the $m=1/2$ incompressible state is explicitly interaction-induced.
Indeed, although the charge excitation gap $\Delta_{\rm ch, 1/2}$ 
is derived in the strongly interacting limit $K\gg 1$, Eq.~(\ref{Delta-12}) 
gives a correct noninteracting limit $\Delta_{\rm ch, 1/2}=0$ 
expected from the single particle Bloch theory. 
Formally, when $K=1$, both flavor 
and charge gaps are zero since 
the last two terms of the potential (\ref{bar-L-12}) vanish ($\kappa\to 0$).

The weak coupling estimate (\ref{Delta-12}) is valid when the flavor 
soliton size (\ref{l12}) is large,
\be \label{def-weak-12}
\lfl_{,1/2} \gg \lsaw \,, 
\ee
according to the condition (\ref{weak}) above. 
Practically, due to the large bandwidth $D$, 
Eq.~(\ref{def-weak-12}) requires rather small bare gap $\Delta_0$.
For typical parameter values,
$\lsaw \sim 0.1~\mu$m and $\Delta_0 = 10$\,meV,
the soliton scale defined in Eq.~(\ref{l12})
is small compared to the potential period $\lsaw$,
$\lfl_{,1/2}\sim 10\,$nm$\,<\lsaw$,  
and the condition (\ref{def-weak-12}) does not hold.
In this case the excitation gap is given by the strong coupling limit
expression (\ref{minigaps-ren-4})
(Section~\ref{sec:strong} below).

However, the result (\ref{Delta-12}) is applicable for realistic parameter
values, whenever the flavor soliton size $\lfl_{,1/2}$ 
becomes large: Either when one takes the tube that is almost metallic,
$\Delta_0\sim 1$\,meV, in which case $\lfl_{,1/2}\sim 1\,\mu{\rm m} >\lsaw$,  
or close to certain potential amplitude values $A^*$ that correspond to 
zeroes of $v_{1/2}(2\tilde A)$ (Fig.~\ref{fig:v12}).

In the special case when the potential amplitude 
is such that the coupling $g_{1/2}$ vanishes, $v_{1/2}(2\tilde A) = 0$, 
the charged excitation is {\it gapless} 
[similar to the integer $m$ case considered above, 
where the gaps (\ref{minigaps-ren})
vanish at the zeroes of the Bessel functions],
but {\it the flavor sector remains gapped}. Its gap $\Delta_{\rm fl}$ 
can be estimated from the effective O(6)$\simeq$SU(4) Gross-Neveu Lagrangian
of the form (\ref{L-GN}) that is given by the last term of the 
Lagrangian (\ref{bar-L-12}), where now the 
coupling $g_{\rm GN} \propto u_{1/2}(2\tilde A^*) \neq 0$ (Fig.~\ref{fig:v12}).
We stress that the resulting Gross-Neveu coupling is a functional 
of the applied potential and thus, to an extent, can be controlled externally. 
Physically, the origin of the flavor gap corresponds to the 
additional cost of destroying the flavor order in a commensurate state
by adding an extra electron of a particular flavor. 
Naturally, $\Delta_{\rm fl}\sim K^{-1/2}\Delta_{\rm ch}$ 
is controlled by the strength of the exchange interaction.

\section{Strong coupling limit}
\label{sec:strong}

\subsection{The Lagrangian for the charge sector}
\label{sec:L-charge}

Under the condition (\ref{strong}), the Wigner crystal saddle point
(\ref{F-strong}) combined with integrating over the flavor fluctuations 
yields the following Lagrangian for the charged mode
[cf. Eq.~(\ref{Lcharge})]:
\bea 
\nonumber
\L_{\rm charge}[\theta^0] = {\hbar v\over \pi} 
\lf 
\frac1{2v^2}(\partial_t\theta^0)^2 
- \frac{K}2 (\partial_x\theta^0)^2
\right. \\ \left.
- g f_1 \cos (4\theta^0 + \mtot\ksaw x - 8\tilde A \sin\ksaw x) 
\vphantom{\int}\rf  .
\label{Lcharge-ext}
\eea
Here the density (\ref{rho-alpha}) is related to the chemical potential 
$\mu$ by Eq.~(\ref{mu-nu}), $8\tilde\mu\to \mtot$, 
and the coupling constant $g$ 
is given by its renormalized value (\ref{g-ren}).

The Lagrangian (\ref{Lcharge-ext}) describes the effective theory 
of a single mode in the presence of the external potential 
that is adiabatic on the scale $\lfl$.
This is a result of integrating out the three neutral modes, whose effect 
has been to produce the Wigner crystal saddle point [Sec.~\ref{sec:switch}]
and to renornalize the coupling $g$.
As a result, the number of degrees of freedom in the theory (\ref{Lcharge-ext}) 
is significantly reduced. 
This simplification is justified under the following assumptions:
(i) separation of charge and flavor energy scales 
$\Delta_{\rm ch}\gg \Delta_{\rm fl}$ 
(for that, strong interactions $K\gg 1$ are crucial); 
(ii) dilute limit $\bar\rho \lfl \ll 1$ that ensures 
exponentially small flavor-flavor correlations.
These conditions imply that the flavor physics is decoupled from the charged one.
In other words, the allowed flavor configurations in a train of composite solitons
that represent electrons at low density enter with equal weights,
which allows us to {\it trace} over them, effectively eliminating them
from the action.
Physically, such a situation corresponds to the limit when
the temperature $T\to 0$ and exchange scale $\Delta_{\rm fl}\to 0$ in such a way
that $\Delta_{\rm fl}/T \to 0$.\cite{CZ,BalentsFiete}

We stress that in the limit (\ref{strong}), 
the renormalized coupling $g\simeq 1/\lfl^2$ in the Lagrangian (\ref{Lcharge-ext})
does {\it not} depend on the density and on the external potential.
This is contrary to the weak coupling case of Section~\ref{sec:weak}
where the renormalization of nonlinear couplings was sensitive to $m$ and $A$.
The reason for this sensitivity in the weak coupling limit 
is in the flavor soliton size $\lfl$ 
(that controls the scale over which the fluctuations are accumulated)
extending over a few potential periods.
Naturally, in that case the form of $U(x)$ 
must influence the RG flow.

In the present case, $\lfl \ll \lsaw$, and the RG flow
produces the effective coupling $g$ [Eq.~(\ref{g-ren})] 
on the scale $a < l < \lfl$ that appears
microscopic for the external potential (\ref{SAW}).
In other words, the adiabatic (on the scales $l\leq \lfl$)
external potential has the effect of the chemical potential that can
at best affect the local charge density, but not the RG flow.

\subsection{Refermionization}
\label{sec:referm}

We now complete our effective single-mode description by 
refermionizing the theory (\ref{Lcharge-ext}).
For that we first  present this Lagrangian in the canonical form 
by rescaling the charge mode:
\be \label{def-Theta}
\Theta=2\theta^0 = \ts\sum_{\alpha=1}^4 \Theta_{\alpha} \,.
\ee
The field $\Theta$ is the canonical displacement field for the total density,
\be \label{rho-Theta}
\rho=\frac1{\pi} \partial_x \Theta \,.
\ee
To preserve correct commutation relations,
the conjugate momentum $\Pi_{\Theta}$ for the field (\ref{def-Theta}) should be 
half of that for the field $\theta^0$,
\be \label{Pi-theta}
\Pi_{\Theta}=\ts{\frac12} \Pi_{\theta^0} = \frac14 \sum_{\alpha=1}^4 \Pi_{\alpha} \,,
\ee
with $\Pi_{\alpha}$ defined in Eq.~(\ref{Pi-alpha}).
Changing variables in the Lagrangian (\ref{Lcharge-ext})
according to (\ref{def-Theta}) and (\ref{Pi-theta})
we obtain the effective Lagrangian for the mode $\Theta$,
\be
\matrix{
\L_{\rm eff}[\Theta] = \ds{\hbar v'\over\pi} 
\lf
\frac1{2v'^2} \lp \partial_t  \Theta\rp^2 
- \frac{K'}2\, \lp\partial_{x}\Theta\rp^2 
\right. \cr \left.
- \frac14 f_1 g \cos(2\Theta+\mtot\ksaw x - 2\tilde A'\sin\ksaw x)
\vphantom{\ds{\int {1\over v^2}}}\rf 
}
\label{L-Theta}
\ee
with the redefined parameters 
\be \label{v'K'}
v' \equiv 4v \,, \quad  K' \equiv \ts{\frac1{16}}K \,, \quad 
\tilde A' = 4\tilde A \equiv \ds{A \over K'\hbar \ksaw v'}  \,.
\ee
The rescaling (\ref{v'K'}) has the following meaning.
The velocity quadrupling simply states that we count incoming fermions
regardless of their original flavors. 
The rescaling of the charge stiffness can be understood by 
noting that, with the same accuracy  
that has allowed us to discard the flavor modes [$K\gg 1$, see above],
\be \label{K'}
K'\approx 1 + {(K-1) \over 16} \equiv 1 + \nu' V(q) \,,
\quad \nu'={1 \over \pi\hbar v'}  \,,
\ee
which is by definition the charge stiffness for the spinless Dirac fermions
of velocity $v'$ and density of states $\nu'$
[cf. Eq.~(\ref{K})]. Finally, the external fields $U$ and $\mu$
are not rescaled [cf. Eqs.~(\ref{tilde-A}) and (\ref{LF})].

Refermionizing the Lagrangian (\ref{L-Theta}) by introducing 
the Dirac spinors $\Psi\simeq (2\pi a')^{-1/2} e^{i\Theta}$, we formally obtain the 
effective Hamiltonian 
\be 
\matrix{
\H_{\rm eff}[\Psi] = \ds{\int}\! dx \, 
\Psi^+ \lf -i\hbar v'\sigma_3 \partial_x  + \Delta' \sigma_1 
+ U(x) \right. \cr \left.
- \mu \rf \Psi 
+ \ds{\frac12}\sum_k\rho_{-k}V(k)\rho_{k}
}
\label{H-strong}
\ee
for the fictitious spinless Dirac fermions.
These are the NT electrons 
averaged (traced) over their SU(4) flavor configurations 
with equal weights.
Naturally, the total fermion number density (\ref{rho}) in the new variables
\be \label{rho-Psi}
\rho(x) = \Psi^+\Psi \,.
\ee
The effective gap $\Delta'$ in Eq.~(\ref{H-strong}) 
is chosen in such a way that it corresponds
to the renormalized coupling (\ref{g-ren}) entering the Lagrangian 
(\ref{L-Theta}),
\be \label{g-Delta'}
{f_1\over 4} g\simeq {\Delta' \over \hbar v' a'} \,, \quad a'\simeq \lfl 
\ee
similarly to the definition (\ref{def-lambda0}).
In Eq.~(\ref{g-Delta'}) the new length scale cutoff $a'$ is assumed since
the present approach is valid only at the length scales $l > \lfl$
beyond the correlation length of the neutral sector.\cite{log}
Eq.~(\ref{g-Delta'}) together with Eqs.~(\ref{g-ren}) and 
(\ref{l0-self-consist}) estimates
\be \label{Delta'}
\Delta' \simeq {\hbar v' \over \lfl} \simeq D^{1/5} \Delta_0^{4/5} \,.
\ee

To summarize, the NT electron dynamics in the dilute limit (\ref{strong})
is described by the effective Hamiltonian (\ref{H-strong}) of 
spinless Dirac electrons. These fictitious fermions
of the density (\ref{rho}) carry unit charge,
and interact with each other and with external fields
in a standard way.

\subsection{Excitation gaps}

The (charge) excitation gaps can be now estimated from 
the effective Lagrangian (\ref{L-Theta}) or from the Hamiltonian (\ref{H-strong}).
Below we assume sufficiently long range interaction
\be \label{deloc}
\lch > \lsaw
\ee 
in which case the charge excitation is extended over several potential minima,
and the single-mode phase soliton approach [Sec.~\ref{sec:phasesoliton}]
applies, with the system (\ref{L-Theta}) of the form (\ref{H1})
with $K\to K'$, $b\to 2\tilde A'$, $\beta=2$, and with
the period ratio defined in terms of the {\it total} density, $m\to\mtot$.

The excitation gaps are then obtained in a standard way:
According to Sec.~\ref{sec:phasesoliton}, 
when the total density $\mtot$ is integer,
one averages the potential term in Eq.~(\ref{L-Theta})
over the period obtaining the effective coupling
\be \label{g-strong}
\bar g = g J_{\mtot}(2\tilde A') \,,
\ee
then integrates over the fluctuations of the $\Theta$-field 
\be \label{rg-strong}
\bar g(l) = \bar g(a') \lp {l\over a'} \rp^{-\eta'} , 
\quad \eta'= \ds{1\over \sqrt{K'}} \simeq 4\eta \ll 1 \,, 
\ee
with $K'$ given by Eq.~(\ref{K'}) and $\eta$ by Eq.~(\ref{eta}),
finds the corresponding charge soliton scale self-consistently as
\be \label{selfconsist-strong}
\bar g(\lch) \simeq \ds{K'\over \lch^2} \,,
\ee
and obtains the excitation gap $2\bar\Delta_m$, where
\be \label{minigaps-ren-4}
\bar \Delta_{m}\simeq \left| J_{\mtot}(2\tilde A')\right|^{\frac1{2-\eta'}} 
\bar \Delta  \,, \quad \mtot =4m = \pm1, \pm 2, ... \,.
\ee 
Here the charge gap 
\be \label{Delta-strong}
\bar\Delta \simeq  K^{\frac{1-\eta'}{2-\eta'}} \Delta' \,,
\ee
with $\Delta'$ given by Eq.~(\ref{Delta'}), 
and $J_{\mtot}\equiv J_{4m}$ is 
the Bessel function that depends on the total density (\ref{rho-alpha}).

When the density $\mtot=p'/q'$ is a simple fraction,
the charge density in the commensurate configuration is $q'\lsaw-$periodic 
[as shown in Fig.~\ref{fig:soliton-sketch}(a) for $\mtot=1/2$]. 
The gaps then follow from the standard phase soliton approach\cite{Bak}
[Sec.~\ref{sec:phasesoliton}] which we will not repeat here.

\subsection{Classical limit}

\nin
The classical limit of the Hamiltonian (\ref{H-strong}) 
is obtained by discarding the ${\cal O}(\eta')$ quantum fluctuations, and  
coarse-graining beyond the length scale $\lfl$, effectively treating 
electrons as {\it distinguishable point particles} 
[on the scale $l > \lfl$] 
that interact with each other via the Coulomb potential (\ref{V}).
The classical energy of this system 
\begin{eqnarray}
\matrix{\ds
E_{\rm cl} = 
\sum
\lb \Delta' + U(x_i) \rb 
\ + \ 
\sum
\lb \Delta' - U(y_j) \rb 
\cr \ds
+ \sum_{i>i'} V(x_i-x_{i'})
+ \sum_{j>j'} V(y_j-y_{j'})
- \sum_{i,j} V(x_i-y_j) . \quad
}
\label{hamiltonian-cl}
\end{eqnarray}
Here the indices $i,i'$ and $j,j'$ run over electrons (with positions $x$)
and holes ($y$), 
correspondingly, in the minima and in the maxima of the potential (\ref{SAW}).
The incompressible states are parametrized by the pairs $(n_e,n_h)$ of 
rational numbers of electrons and holes per period $\lsaw$
according to Eq.~(\ref{ne-nh})
[cf. Section~\ref{sec:intro} and Fig.~\ref{fig:soliton-sketch}].

The connection between the classical limit (\ref{hamiltonian-cl}) 
of the effective Hamiltonian (\ref{H-strong}), and 
the original charge mode Lagrangian (\ref{L-Theta}) 
follows from representing the coordinates of electrons 
by
\be \label{xj}
x_j = x_j^{(0)} + \phi_j \,, 
\ee
where $\phi_j$ are the displacements from the ideal positions 
\be \label{xj0}
x_j^{(0)} = {j\lsaw \over \mtot} \,, \quad j=1, 2, ... 
\ee
in the Wigner crystal in the absence of the external potential.
[For simplicity, we are not considering the holes in the maxima of $U(x)$, 
which can be treated analogously.]
In the continual limit $\phi_j \equiv \phi(x_j) \approx \phi(x)$ 
the charge density is 
\be \label{rho-xj}
\rho(x) 
\approx {\mtot\partial_x \phi \over \lsaw} \,, 
\ee
describing the change of $\phi$ by 
$\lsaw/\mtot$ when one extra particle is added. 
The third sum in Eq.~(\ref{hamiltonian-cl}) gives the interaction energy 
between the electrons. Substituting the coordinates (\ref{xj}) into this 
sum and expanding it up to the second order in  
$\phi_i-\phi_j \approx  \lp x_i^{(0)}-x_j^{(0)} \rp \partial_x \phi$, 
we obtain the gradient term for the displacement mode energy,
\be \label{Ekin}
\frac12 \int \! dx\, \lp 2e^2 \ln \tilde N \rp \rho^2(x) \equiv  
{\hbar v' \over \pi} \int \! dx\, {\tilde K' (\partial_x \Theta)^2\over 2}
\,, 
\ee
where 
\be
\Theta = {\pi \mtot \over \lsaw}\phi(x) 
\ee
is the net charge mode (\ref{def-Theta})
defined in accord with the expression
(\ref{rho-Theta}) for the total charge density, 
and the charge stiffness 
\be
\tilde K' = {1\over \pi \hbar v'} \cdot 2e^2 \ln \tilde N  \simeq K'-1 \approx K' \,.
\ee
Here the stiffness $K'$ is defined in Eq.~(\ref{K'}),
and the argument $\tilde N$ of the Coulomb logarithm is found 
self-consistently as a number of electrons 
whose positions are altered in a charged phase-soliton excitation.
The expression (\ref{Ekin}) is the classical limit of the second term 
of the Lagrangian (\ref{L-Theta}) with a meaning of the Coulomb
interaction between the quasi-classical electrons.
The nonlinear term of the Lagrangian (\ref{L-Theta}) gives the bare energy cost 
$\Delta' \cos \lp \mtot \ksaw \delta\phi\rp = \Delta'$
of adding a particle above the gap (modulo interactions), 
with $\delta\phi=\lsaw/\mtot$ according to Eq.~(\ref{rho-xj}). This term corresponds
to the constant term in the classical energy (\ref{hamiltonian-cl}),
$\sum_{x_i}\Delta'$. 
The interaction with the potential,  $\sum_{x_i} U(x_i)$,
can be written in the usual form (\ref{Hext}) and added to the argument 
of the cosine via the gauge transformation (\ref{theta-gauge}).

\subsection{Discussion}

\nin
The main result of this Section is the single-mode effective Hamiltonian
(\ref{H-strong}) that has allowed us to map the problem of the interacting
fermions with the four flavors onto that of a single flavor and to utilize
the standard phase soliton approach for the single mode.

The physical meaning of the present treatment is as follows.
In the noninteracting case, fermions of the same flavor 
avoid each other due to the Pauli principle.
The ground state wave function is then given by 
a product of the four Slater determinants, one for each flavor.
However, when the repulsion between the fermions is strong,
fermions of {\it all} the flavors avoid each other in a similar way,
and the ground state wave function is a Slater determinant of 
a four-fold size.\cite{Ogata} 
This is manifest in the $\mtot=4m-$dependence 
of the excitation gaps  (\ref{minigaps-ren-4}).

The original SU(4) flavor symmetry of the problem
becomes manifest on the level of renormalization,
namely in the particular scaling law of $4/5$ 
of the renormalized gap (\ref{Delta'}) produced by the renormalization
group flow of the flavor sector on the length scales $a < l < \lfl$.

Let us compare the scaling laws in  Eqs.~(\ref{minigaps-ren-4})
and (\ref{Delta-strong}) with those obtained in the weak-coupling limit. 
In the classical limit 
the charge excitation gap (\ref{Delta-strong}) for the stand-alone nanotube
has the same form as that obtained in Sec.~\ref{sec:fluct} above,
Eq.~(\ref{Delta-Kinf}):
\be
\bar\Delta = \Delta \quad {\rm when} \quad K\to \infty \,.
\ee
However, one notes that at finite $K$, the power law exponents 
in the expressions (\ref{Delta-strong})
and (\ref{minigaps-ren-4}) are {\it different} from those in 
Eqs.~(\ref{Delta-K}) and (\ref{minigaps-ren}).
This is not surprising since the theory (\ref{Lcharge-ext}) is different from
the original model (\ref{L'}). Whereas in the latter the exchange is important,
in the former the flavor sector is traced over under the assumptions
specified above in Sec.~\ref{sec:L-charge};
different theories indeed yield different scaling laws.
A similar subtlety in taking the order of limits of vanishing temperature and 
exchange 
was observed in the recent calculation of correlation 
functions \cite{CZ} for spin$-1/2$ fermions, 
and was qualitatively explained in Ref.~\onlinecite{BalentsFiete}.
In the present case, the same phenomenon is manifest in the scaling behavior 
of the excitation energy.

We underline that the external potential (\ref{SAW}), by 
adding an extra length scale $\lsaw$,
naturally distinguishes the regimes $\lfl > \lsaw$ and  $\lfl < \lsaw$
in which the flavor physics is important and unimportant, correspondingly.
Accordingly, the dependence on the 
{\it parameters of the potential} in these two regimes is also 
qualitatively different,
as one may see by comparing {\it e.g.} Eqs.~(\ref{minigaps-ren}) and 
(\ref{minigaps-ren-4}) {\it even in the limit $K\to \infty$}.

\section{Phase diagram}
\label{sec:weak-tunneling}

\begin{figure}[b]
\includegraphics[width=3.5in]{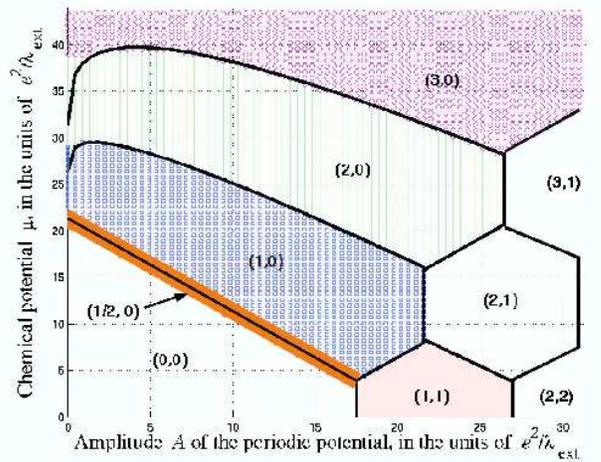}
\caption[]{(Color online)
Phase diagram for the nanotube,
$\Delta'=6\pi e^2/\lsaw$, $\ln(L/\lsaw)=2.5$,
$e^2/\lsaw = 1.44$ meV for $\lsaw=1\mu$m.
Shaded regions are the result of the numerical minimization of Eq.~(\ref{Enn-mun})
(see text), 
black lines are approximations obtained from Eq.~(\ref{min-energies})
with $V_n$ given by Eq.~(\ref{Vn-local}).
}
\label{fig:sawphdiag}
\end{figure}

\nin
Below we draw the phase diagram for the classical system (\ref{hamiltonian-cl})
in the $A,\mu$ plane, Fig.~\ref{fig:sawphdiag}, where $A$ is the potential 
amplitude in Eq.~(\ref{SAW}).
Here we assume that the screening length  is larger 
than the length $L$ of the system, 
\be \label{def-weak-tunneling}
l_s\sim \lch > L  \equiv N \lsaw 
\ee
(effectively meaning that $\lch = \infty$), 
thus the charged phase-soliton optimization of an excitation 
(described in Sec.~\ref{sec:phasesoliton}) does not take place.
In the limit (\ref{def-weak-tunneling}), 
by raising the chemical potential one adds electrons (holes) 
to the sequence of ``quantum dots'' ( ``anti-dots'')
evenly over all the system, changing the commensurate density $\mtot$
from one rational number to another.
Each region in Fig.~\ref{fig:sawphdiag} 
corresponds to a particular commensurate phase $(n_e,n_h)$ 
(as described in Section~\ref{sec:intro}), with density 
(\ref{ne-nh}).
Borders separating regions with integer $(n_e, n_h)$ 
are comprised of fractional-density states, such as 
the domain $(1/2, 0)$ between $(0, 0)$ and $(1, 0)$ 
[charge state sketched in Fig.~\ref{fig:soliton-sketch}(a)].
At zero temperature, all the commensurate charge states 
(rational $\mtot=p'/q'$) 
are incompressible, with a spectrum being a devil's staircase.
At finite temperature gaps for sufficiently large denominator $q'$
will be washed out. The underlying Dirac symmetry makes the 
phase diagram symmetric with respect to 
$\mu \leftrightarrow -\mu, \ n_e \leftrightarrow n_h$, so that only the 
$\mu>0$ part is shown.

The phase diagram is obtained by minimizing the energy functional 
\be \label{Enn-mun}
E_{n_e,n_h}(A) - \mu(n_e-n_h) \,, \quad 
E_{n_e,n_h} = \ts{\frac1N} E_{\rm cl}(x_i,y_j) 
\ee
[$E_{\rm cl}$ given by Eq.~(\ref{hamiltonian-cl})],
with respect to positions $x_i$ and $y_i$ of electrons and holes
in the following way. 
For large size $N$,  we neglect finite size effects and utilize 
translational invariance. In this case, in a commensurate state, 
the optimal positions 
of electrons and holes relative to each potential minimum (maximum) are the same
for each period.
The result is the family of system's charging energy values $E_{n_e,n_h}(A)$
per period $\lsaw$, such as  
\bea
E_{10} &=& \Delta' - A + {e^2\over \lsaw}\sum_{n=1}^{N/2} {1\over n} \,,\\
E_{11} &=& 2E_{10} - {2e^2\over \lsaw} 
\sum_{n=1}^{N/2} {1\over n - \frac12} \,,
\eea
\bea
\nonumber
E_{20} = \mathop{{\rm min}}_{\delta x} \lf \vphantom{{{{{{X^X}^X}^X}^X}^X}^X}
2\Delta' - 2A\cos \ksaw \delta x + {e^2\over 2\delta x} 
\right. \\ \left.  
+ {e^2\over \lsaw}
\sum_{n=1}^{N/2} 
\lp {2\over n} + {1\over n+{2\delta x/\lsaw}} + {1\over n-{2\delta x/\lsaw}}
\rp
\rf \quad
\eea
[here $\pm\delta x$ is the electron coordinate relative to the minimum of $U(x)$],
and so forth. While minimizing the energy $E_{n_e,n_h}$ with respect to 
the positions of electrons and holes within each potential period, unphysical 
configurations of electrons on top of the holes are excluded by demanding
that their minimum separation be $e^2/\Delta'$. The latter condition takes
into account exciton binding energy in the Dirac system.

\begin{figure}[t]
\includegraphics[width=3.4in,height=2.5in]{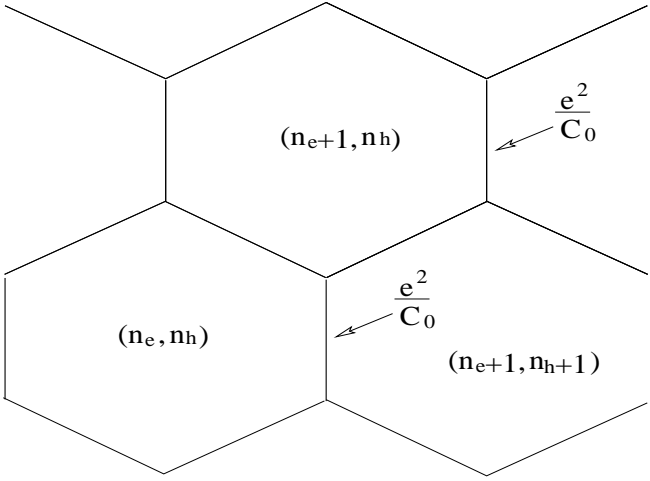}
\caption[]{Phase diagram in the $(A, \mu)$ plane 
according to the model (\ref{min-energies}).
The minimum width of the gap for the incompressible state with any integer
density $n_e-n_h$ is given by the NT charging energy (\ref{gap-min}).
In the limit  $n_e, n_h \gg 1$ all the regions of the 
phase diagram are identical, with their period in $A$
given by Eq.~(\ref{gap-width}).

}
\label{fig:phdiag-detail}
\end{figure}

The functional $E_{n_e,n_h}(A)$ can be further approximated  
by minimizing the interaction energy between the charges only in the 
{\it same} potential minimum (or maximum), treating the rest of the system
in a mean-field way: 
\be 
\label{min-energies}
E_{n_e, n_h} \approx (n_e+n_h)(\Delta'-A) 
+{(n_e-n_h)^2 e^2 \over 2C_0}
+ V_{n_e} + V_{n_h} \,.
\ee
The first term in (\ref{min-energies}) is the
energy of $n_e$ electrons placed
into each minimum and $n_h$ holes into each maximum of $U(x)$. 
It corresponds to the
first two terms of (\ref{hamiltonian-cl}).
The second term in (\ref{min-energies}) is the interaction energy
of the electrons and holes located in the extrema of $U(x)$. Here 
\be \label{def-C0}
C_0 \simeq {\lsaw\over 2\ln(L/\lsaw)} 
\ee
is the NT capacitance per period $\lsaw$.
Finally, $V_{n}$ in (\ref{min-energies}) is the interaction energy of
$n$ electrons (or $n$ holes) minimized with respect to their positions
inside the corresponding potential well of the  
periodic potential (\ref{SAW}). 
The Dirac symmetry yields $E_{n_e, n_h}=E_{n_h, n_e}$.
Minimization of Eq.~(\ref{Enn-mun}) using the approximation (\ref{min-energies}) 
yields the phase diagram sketched in Fig.~\ref{fig:phdiag-detail}.
In the limit (\ref{def-weak-tunneling}),
excitation gaps corresponding to the incompressible states with the total 
density $n=n_e-n_h=(n_e+1)-(n_h+1)=...=(n_e+s)-(n_h+s)$ 
oscillate but do not vanish. Their minimum value 
\be \label{gap-min}
\delta \mu_{n_e, n_h}^{\rm min} = {e^2\over C_0}
\ee
is determined by the NT charging energy. 
The regions $(n_e, n_h)$ and $(n_e+1, n_h+1)$ of the phase diagram 
are separated by the vertical 
lines of fixed $A$, with its value implicitly determined from 
\be \label{A-gap}
A=\Delta'
+ \ts{\frac12} \lf V_{n_e+1}(A)-V_{n_e}(A) + V_{n_h+1}(A)-V_{n_h}(A)\rf .
\ee
Away from the values (\ref{A-gap}) the gap increases $\propto A$.

The mean field approximation (\ref{min-energies}) produces fairly accurate borders
between the domains of the phase diagram in Fig.~\ref{fig:sawphdiag}.
Below we consider the quantum-dot 
addition energies $V_{n_e}$ for the small and large $n_e$.

\subsection{Small density $n_e, n_h \sim 1$}

First consider the situation when
the Coulomb interaction over a period is small, ${e^2/\lsaw} \ll A$,
such that $n_e$ electrons are grouped close to the minima of the potential  
that can be approximated by a piecewise-quadratic polynomial: 
\be \label{Uapprox}
U(x) \approx  -A + {\rm min}_{n\leq N} \frac{A\ksaw^2}2 
\lp x-(n+\ts{\frac12})\lsaw \rp^2 .
\ee
Minimizing the Coulomb energy of $n_e$ charges in a single minimum of the potential
(\ref{Uapprox}), one obtains the values 
\bea
\label{Vn-local}
\matrix{
V_0=V_1=0 \,, \cr
V_2=3\,(\pi/2)^{2/3}\lp{e^2\over \lsaw}\rp^{2/3} A^{1/3} \,, \cr
V_3=5^{2/3} V_2(A) \,,  \, ... \,. 
}
\eea
The power law correction $\sim A^{1/3}$ due to the above expressions 
is observed in Fig.~\ref{fig:sawphdiag} for $A\sim \Delta'$
as a deviation from the straight lines that separate different 
regions of the phase diagram.

When, on the other hand, the periodic potential is a small 
perturbation that shifts electrons from their
equidistant position in a Wigner crystal ($A\ll e^2/\lsaw$),
the perturbation theory in $A$ yields analytic behavior 
of the gap widths on $A$.

\subsection{Large density $n_e, n_h \gg 1$}

\nin
At large $n_e, n_h \gg 1$ the model (\ref{min-energies}) 
can be simplified by using the continuous Thomas-Fermi
description for the density of the classical electrons (holes) 
inside each potential minimum (maximum).  
To find $V_{n_e}$, we approximate the local charge density as 
\be \label{q}
\rho(x) \approx \rho_{\rm TF} = -{\pi n_e \over \lsaw} \cos \ksaw x \,,
\ee
that mimics the external potential profile (\ref{SAW}).
The density (\ref{q}) is normalized to 
$n_e=\int_{\lsaw/4}^{3\lsaw/4} \rho_{\rm TF}(x) dx$. 
The interaction energy $V_{n_e}$ or $V_{n_h}$ inside each ``quantum dot'' can 
be estimated as 
\be \label{V-C1}
V_{n_e} \simeq {e^2\over 2} \int_{\lsaw/4}^{3\lsaw/4} \! dx dx'\, 
{\rho_{\rm TF}(x)\rho_{\rm TF}(x') \over |x-x'|} 
\equiv {n_e^2 e^2\over 2C_1}  \,,
\ee
where the ``dot capacitance'' 
\be \label{def-C1}
C_1 \simeq {\lsaw \over 2\pi \ln {\lsaw\over a'}} 
\ee
is independent of the potential amplitude $A$.
The log singularity in the integral in Eq.~(\ref{V-C1})
is cut off on the scale $a'\sim \lfl$ of the order of the flavor soliton size,
below which the quasiclassical description breaks down
(see Section~\ref{sec:strong}).
The $A$--dependence of $C_1$ would appear as a correction to $V_{n_e}$ 
with $A$--dependent integration limits in Eq.~(\ref{V-C1}).
The case $V_{n_h}$ 
of the holes in the potential maxima (``anti-dots'') is analogous.
Eq.~(\ref{A-gap}) then yields the potential amplitude values 
that separate the configurations $(n_e, n_h)$ and $(n_e+1, n_h+1)$:
\be \label{As}
A_{(n_e, n_h) \to (n_e+1, n_h+1)} = \Delta' + {e^2\over 2C_1} 
(n_e+n_h+1) \,. 
\ee
In this limit all the honeycomb regions in Fig.~\ref{fig:phdiag-detail}
are identical with their size in $A$ being 
\be \label{gap-width}
\delta A_{n_e, n_h} = {e^2 \over C_1} \,.
\ee
Qualitatively, the regular honeycomb structure of Fig.~\ref{fig:phdiag-detail} 
appears already for moderate amplitudes $A$
(Fig.~\ref{fig:sawphdiag}). The borders between the regions 
of the diagram in the limit 
$A > e^2/\lsaw, \Delta'$, are approximately linear, dominated by
the linear dependence of $E_{n_e, n_h}$ on $A$ that stems 
from the first term of Eq.~(\ref{min-energies}).

\subsection{Crossover with the case $\lch < L$}

\nin
We will now show that the asymptotic values of the positions (\ref{As})
of the gap minima derived above for large $n_e$ and $n_h$ 
coincide with those obtained in Section~\ref{sec:strong}
utilizing the phase soliton method in an infinite system. For this crossover
we assume $\lch\sim l_s \sim \lsaw$.

Consider the positions of the minima for excitation gaps (\ref{minigaps-ren-4}) 
(as a function of the potential amplitude)
\be
\tilde A' = 4\tilde A = {4A \over K \epsilon_0} \approx
{\pi\over 2} \cdot {A\over e^2/C_1} \,, 
\ee
where we took the value of the charge stiffness (\ref{K})
at the momentum scale $\lsaw^{-1}$ and substituted $K\to K-1$ at $K\gg 1$. 
Utilizing the asymptotic behavior of the $n-$th zero of the Bessel function
$J_{\mtot}$,
\be
2\tilde {A'}_{\mtot}^{(n)} \simeq {\rm const}\, + {\pi \mtot \over 2} + \pi n \,,
\quad n={\rm min\, } \{ n_e, n_h \} \,,
\ee
we indeed obtain the values 
$A\simeq (n_e+n_h)e^2/2C_1$ corresponding to gap minima
in accord with Eq.~(\ref{As}).

\section{Experimental means to probe incompressible states}
\label{sec:current}

\subsection{Conductance measurements}

Excitation gaps corresponding to the incompressible states 
can be detected in transport measurements. 
When, by means of varying the gate voltage, 
the Fermi level is tuned to be in the gap, the zero-bias 
conductance $\sigma$ across the nanotube
vanishes. At finite temperature $T$, $\sigma$ will have the activated form
\be \label{sigma-act}
\sigma \sim e^{-T_{\rm act}/T} \,,
\ee
with the activation energy $T_{\rm act}=\Delta_m$ 
equal to half of the excitation gap corresponding 
to a particular incompressible state. 
Physically, Eq.~(\ref{sigma-act})
corresponds to the transport by means of thermally-excited charge solitons that 
can carry electric current through the nanotube. 
(A similar situation has been recently considered in the context of 
transport in granular arrays.\cite{AGK})
The values and positions of the gaps 
can be also revealed by measuring the differential conductance at finite bias 
with varying gate voltage.
In this case, peaks in the differential conductance 
as a function of the applied bias will mark the positions of the excitation gaps.

\subsection{Adiabatic charge pumping}

A more challenging possibility is to realize the 
Thouless pump\cite{Thouless} in a nanotube. 
Such a setup requires an adiabatically moving 
periodic potential that could be created {\it e.g.} by coupling to   
a surface acoustic wave (SAW),\cite{Talyanskii'01}
or by sequentially modulated gate voltages on the array of underlying gates.
When the chemical potential  is inside the $m$-th minigap $\Delta_m$, 
the adiabatically moving periodic potential with frequency $f$
induces {\it quantized current} \cite{Talyanskii'01} 
\be \label{j-quantized}
j = \mtot ef \,, \quad \mtot=4m \,.
\ee

In such a setting, novel fractional-$m$ incompressible states 
considered in this work will manifest themselves in
the adiabatic current (\ref{j-quantized}) quantized in 
the corresponding {\it fractions} of $4ef$.\cite{NT-devil}
This result can be understood by invoking the topological invariant property
of the Thouless current.\cite{Thouless,niu-thouless}
The expression (\ref{j-quantized}) is trivial in a semiclassical limit,
with a meaning of transporting on average $\mtot$ electrons per cycle
in a conveyer-belt fashion.
By staying inside the gap and adiabatically changing
the parameters of the system, the current (\ref{j-quantized}) 
remains invariant and hence it is valid in the fully quantum-mechanical case.

Operation of the charge pump requires adiabaticity
\be \label{adiabaticity}
k_B T , \ hf \ll \Delta_m \,.
\ee 
Since the energy scale for the minigaps is set either by the 
gap $2\Delta_0$ at half-filling, or by the strength of the Coulomb interactions
between electrons separated by $\sim\lsaw$,
the typical minigap values $\Delta_m$ are in the meV range, and the 
adiabaticity condition (\ref{adiabaticity}) is realistic. 
The feasibility of the Thouless pump in the NT-SAW setup is further 
corroborated by  recent pumping experiments involving SAWs.
In particular, in the pumping of electrons between the two  
2D electron gases through a pinched point contact \cite{Shilton} 
the achieved quality of current quantization is close to 
metrological.\cite{niu'90,current-standard}
Recently the SAW-assisted pumping has been demonstrated 
through the laterally defined quantum dot,\cite{screw}
as well as through the semiconducting nanotube whose working 
length $L$ matched the SAW period, $L=\lsaw$.\cite{Ebbecke'04,Leek}

We contrast the non-dissipative current (\ref{j-quantized}) {\it on the plateau}
with the dissipative non-quantized current away from commensuration
considered in Ref.~\onlinecite{Niu} for arbitrary quantum wire.
Such a non-quantized current is pumped when no gap opens
(for incommensurate densities or for commensurate densities with the
interaction below criticality) and is characterized by the 
interaction-dependent critical exponents.\cite{Niu}

A practical realization of the proposed pumping setup 
can become a first implementation of the Thouless transport. 
Besides being instrumental in studying electron interactions in a nanotube
(by detecting and measuring fractional$-m$ minigaps that arise solely due to 
interactions), it could realize the ``conveyer belt'' for electrons
with a possibility of pumping current quantized in fractions 
of the unit charge per cycle.
In other words, such a setup would make the first example 
of the charge pump operating at the {\it fraction of the base frequency}.
This setup could allow one to study in detail 
the electron correlations and interactions in  
nanotubes, with a variety of controllable parameters at hand, 
such as the shape and frequency of the external potential,
nanotube gap $\Delta_0$ (that can be modified externally),
and the gate voltage. By exploring
the phase diagram one can study effects of Wigner crystallization,
quantum commensurate-incommensurate transitions, 
and the Tomonaga-Luttinger correlations.

Finally, we note that the described setup can be utilized
to adiabatically transport low-energy strongly correlated SU(4)
{\it flavor} states [{\it e.g.} those for $m=1/2$ obeying the effective 
Gross-Neveu Lagrangian of the form (\ref{L-GN}) described in Sec.~\ref{sec:weak}] 
over a macroscopic distance, 
since the coupling to the adiabatically moving external potential is SU(4) invariant
and thus it does not destroy spin or flavor correlations.
This may become useful in the context of  
solid state implementations of quantum information processing.
Furthermore, this setup can provide a possibility 
of realizing a quantized spin-polarized pump 
by subjecting the system to magnetic field tuned in such a way 
that only one spin population of electrons is commensurate with the potential and 
participates in current quantization.

\section{Conclusions}
\label{sec:conclusions}

In the present work 
we have shown that coupling of the interacting electrons in a nanotube 
to an external periodic potential is a rich setup to 
study one-dimensional electronic correlations, including the 
crossover between the Luttinger liquid and the Wigner crystal. 
We demonstrated that the external potential 
locks the system into incompressible states. 
The corresponding excitation gaps 
(estimated to be in meV range)
are found by 
adequately treating the curvature of the electronic dispersion in the 
bosonized language, and by further generalizing the 
phase soliton method onto the case of multiple modes.

In the regime when the gaps open due to the Bragg diffraction 
in a multi-flavor Luttinger liquid,
we identified and investigated the novel incompressible 
fractional-density state with 
$m=1/2$ electrons of each flavor per period of the potential.
The phase soliton action derived for this case 
describes the charge excitation, 
and the SU(4)-flavor excitations governed by the O(6) Gross-Neveu model.

In the opposite limit of the pinned Wigner crystal we derived the 
effective single-mode Hamiltonian and found that the 
phase diagram  in the classical limit has a stucture of the devil's staircase.

The interaction-induced incompressible states can be detected in the 
Thouless pump setup that can allow one to
study electron correlations and the transition to the Wigner crystal, 
as well as to realize the quantized charge pump 
that operates at a fraction of the base 
frequency by virtue of electron-electron interactions.

\section*{Acknowledgments}

\nin
It is a pleasure to thank Leonid Levitov for bringing this problem to the author's
attention and for fruitful discussions. 
This work was initiated at the Massachusetts Institute of Technology 
(supported by NSF MRSEC grant DMR 98-08941), 
and completed at Princeton (supported by NSF MRSEC grant DMR 02-13706).


\appendix

\section{Conditions for the tight-binding limit}
\label{app:tb}

\nin
Below we derive the condition (\ref{tb-condition})
for the tight binding limit (\ref{def-strong-nonint})
in the single particle picture.
We consider the two cases depending on the relation between 
the potential amplitude $A$ and the Dirac mass term $\Delta_0$.

The case $A\ll\Delta_0$ corresponds to the ``nonrelativistic limit'' of the
Dirac equation, where all the relevant energies are much smaller than
the gap $\Delta_0$. We define the ``Dirac'' electron mass $M$ via
$\Delta_0 = M v^{2}$, and turn 
to the effective Schr\"odinger description with the Hamiltonian 
\be\label{hamiltonian-S}
\H_{\rm Sch} = - {\hbar^2\over 2M}\, \partial_x^2  + U(x) \,,
\ee
in full analogy with 
the non-relativistic limit in QED. 
The tunneling amplitude
between the adjacent minima of $U(x)$ is proportional to 
$e^{-S|_{A<\Delta_0}/\hbar}$,
where the classical action under the barrier
\be \label{action-sch}
{S|_{A<\Delta_0} } = {4\hbar\sqrt{A \Delta_0} \over \epsilon_0} \,. 
\ee
Therefore both the minibands and tunneling are suppressed 
in the limit $A\ll \Delta_0$ if 
\be \label{weak-tun-sch}
\lp {\Delta_0\over \epsilon_0} \rp^{-1} < {A\over \epsilon_0} 
\ll {\Delta_0 \over \epsilon_0} \,.
\ee

The inherently Dirac regime occurs when $A > \Delta_0$.
In this case electron can 
tunnel between the minima of the potential (\ref{SAW}) sequentially through the
hole part of the spectrum. 
The corresponding tunneling amplitude 
$\sim e^{-2S|_{A>\Delta_0}/\hbar}$, where the 
classical action 
for the particle with energy $E$
under barrier between the electron and hole parts is 
\be \label{def-action-dir}
S|_{A>\Delta_0} = \frac1v \int \! dx \, \sqrt{|[E-U(x)]^2-\Delta_0^2|} \,.
\ee 
Tunneling from the minimum of $U(x)$ at $E=\Delta_0-A$ to the hole part of the
spectrum yields
($\delta\equiv {\Delta_0/A}$)
\bea 
\nonumber
S|_{E=\Delta_0- A} = {2\hbar A \over \epsilon_0} 
\; \lf 
\sqrt{\delta} - \lp 1-\delta\rp 
\ln \frac{1+\sqrt{\delta}}{\sqrt{1-\delta}} 
\rf  
\\ \label{action-dir-series}
=  {4\hbar\Delta_0^{3/2}\over \epsilon_0 A^{1/2}} \; 
\lp 
{1\over 1\cdot 3}  
+ {\delta\over 3\cdot 5} 
+ {\delta^2\over 5\cdot 7} + \; ...
\rp .
\eea
Tunneling from the energy level $|E|\ll A-\Delta_0$ far from the potential bottom
yields the action 
$S|_{|E|\ll A-\Delta_0}\simeq \pi \Delta_0^2/\hbar v {\cal E}$, where 
${\cal E} = |\partial_x U(x)|\simeq \ksaw A$. 
The requirement $S|_{E=\Delta_0-A}> \hbar$ or $S|_{|E|\ll A-\Delta_0}> \hbar$
yields that the single particle bandwidth in the Dirac regime
is exponentially suppressed if 
\be \label{weak-tun-dir}
{\Delta_0\over \epsilon_0} < {A\over \epsilon_0} < 
\lp {\Delta_0\over \epsilon_0}\rp^c ,
\ee
where, correspondingly,
the exponent $c=3$ for tunneling from the potential minimum and 
$c=2$ for tunneling from energy level far from the potential bottom.

Summarizing, Eqs.~(\ref{weak-tun-sch}) and (\ref{weak-tun-dir}) yield
the condition (\ref{tb-condition}) 
for the tight-binding limit (\ref{def-strong-nonint}).


\section{$m=\frac12$ phase soliton action} 
\label{app:12}

\nin
Below we consider the weak coupling limit for the fractional density $m=\frac12$,
characterized by the chemical potential
$\tilde\mu = \frac14$ in accord with Eq.~(\ref{mu-nu}).

Our course of action has been outlined in Section~\ref{sec:int} above.
Technically, it will be more convenient to first work in the basis of the original 
fields $\Theta_{\alpha}$ and later utilize the transformation (\ref{transf})
to obtain the Lagrangian $\L_{1/2}[{\bar\theta^0, \bar\theta^a}]$
in the charge-flavor basis. 
Since the average of the potential energy (\ref{LF}) 
over the {\it two} successive potential periods is zero, 
we decompose  the fields $\Theta_{\alpha}(x)$ 
in a series in the coupling $g_0$ keeping the zeroth and the first orders 
\be \label{bar-theta-alpha}
\Theta_{\alpha}=\bar\Theta_{\alpha} + \Theta^{(1)}_{\alpha} \  ,
\quad \Theta^{(1)}_{\alpha} = {\mathcal O}(g_0) \,.
\ee

\begin{figure}[b]
\includegraphics[width=3.5in]{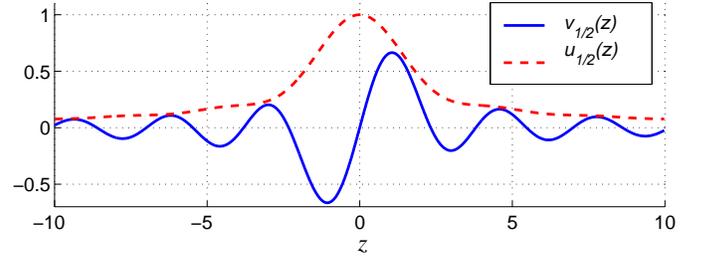}
\caption[]{(Color online)
The functions $v_{1/2}$ and $u_{1/2}$ defined in Eqs.~(\ref{v12}) and (\ref{u12}).
We find that $v_{1/2}(z) < u_{1/2}(z)$ holds for all $z$. Zeroes of $v_{1/2}$:
$z=0, \ \pm 2.33, \ \pm 3.80, \ \pm 5.47, ... $ .
}
\label{fig:v12}
\end{figure}

Consider the Hamiltonian
\bea
\nonumber
\H^{(1/2)}[\Theta_{\alpha}] = {\hbar v\over \pi}
\lf
\ds{{1\over 2}}\, \sum (\partial_{x}\Theta_{\alpha})^2  
+ \ds{{K-1\over 8}}\, \lp\sum \partial_{x}\Theta_{\alpha}\rp^2 
\right. \\ \left.
+ \ds{\frac14} g_0\, \sum \cos ( 2\Theta_{\alpha} 
+ \ts{\frac12} \ksaw x 
- 2\tilde A \sin \ksaw x)
\rf.  \quad 
\label{H-bosonized-gauged-j}
\eea
The Euler-Lagrange equations $\delta \H^{(1/2)}/\delta\Theta_{\alpha}=0$ are
\be
\Theta_{\alpha \ xx}^{(1)}  + {K-1\over 4} S^{(1)}_{xx}  = 
-{g_0\over 2}
\sin ( 2\bar\Theta_{\alpha} + \ts{\frac12}\ksaw x 
- 2\tilde A\sin \ksaw x) , 
\label{eq-on-theta-j}
\ee
where $S  = \sum \Theta_{\alpha}$, and $S_x \equiv \partial_x S$, etc. 
Integrating Eq.~(\ref{eq-on-theta-j}) we obtain
\bea
\label{theta-j-1}
\Theta_{\alpha \ x}^{(1)} &=& {1-K\over 4} S^{(1)}_{x}
+ {g_0 \over \ksaw}\, \tilde\Theta_{\alpha} \,, \\
\label{S-1}
\quad  S^{(1)}_{x} &=& {g_0 \over K\ksaw}\, \sum \tilde\Theta_{\alpha} \,, \\ 
\tilde\Theta_{\alpha} &=& \sum_m {J_m(2\tilde A)\over 1-2m} 
\cos \lp 2\bar \Theta_{\alpha} + (\ts{\frac12}-m)\ksaw x\rp. \quad
\label{tilde-theta-j-1}
\eea
Substituting Eqs.~(\ref{theta-j-1}) and (\ref{tilde-theta-j-1}) 
into the Hamiltonian (\ref{H-bosonized-gauged-j}), after somewhat lengthy but
straightforward algebra the slow mode potential follows:
\bea
\label{bar-V-12}
\nonumber
V_{1/2} &=& {\hbar v g_0'\over 16\pi} 
\lf \vphantom{\sum_{\alpha\neq \alpha'}}
(4-\kappa) v_{1/2} 
\sum_{\alpha} \cos 4\bar\Theta_{\alpha}
\right.  \\ 
\nonumber
&+& \kappa u_{1/2} \sum_{\alpha\neq \alpha'} 
 \cos (2\bar\Theta_{\alpha} - 2\bar\Theta_{\alpha'}) 
\\ 
&-& \left. \kappa v_{1/2} \sum_{\alpha\neq \alpha'}
\cos (2\bar\Theta_{\alpha} + 2\bar\Theta_{\alpha'}) 
\rf. 
\eea
Here 
\bea
g_0' &=&  \lp{g_0 \over \ksaw}\rp^2 
= \lp {\Delta_0 \over \epsilon_0 a}\rp^2 \,,\\
\kappa &=& {K-1\over K} \,,
\eea
and the couplings $v_{1/2}$ and $u_{1/2}$ are defined as 
\bea
\label{v12}
v_{1/2}(z) = \sum_{m=-\infty}^{\infty} {J_m(z)J_{1-m}(z)\over (2m-1)^2} \,,\\
\label{u12}
u_{1/2}(z) = \sum_{m=-\infty}^{\infty} \lp{J_m(z)\over 1-2m}\rp^2 
\eea
with $z$ being a shorthand for $2\tilde A$.
The functions $v_{1/2}(z)$ and $u_{1/2}(z)$ are plotted in Fig.~\ref{fig:v12}.

In the commensurate state the minimum value  
\be
{\rm min\ }V_{1/2}[\bar \Theta_{\alpha}] = 
-{\hbar v g_0'\over 4\pi}
\lf 
4|v_{1/2}(2\tilde A)| 
+ \kappa u_{1/2}(2\tilde A) \rf
\ee
of the potential (\ref{bar-V-12}) corresponds to 
$\{\bar\Theta_{\alpha}\}$ being a permutation of  
a set $\{\phi_1 \ \phi_1 \ \phi_2 \ \phi_2\}$, with 
$\phi_{1,2} = \pm \pi/4$ for $v_{1/2}(2\tilde A)>0$ and
$\phi_{1} = 0, \ \phi_2 = \pi/2$ for $v_{1/2}(2\tilde A)<0$.

We now discuss the obtained commensurate classical state.
The ground state degeneracy in the noninteracting case ($\kappa=0$)
is equal to $2^4$. This follows from the potential 
(\ref{bar-V-12}) in which only the first term is nonzero.
In the presence of interactions ($\kappa > 0$),
the other terms in (\ref{bar-V-12}) reduce this degeneracy 
from 16 to six.  This result could have been foreseen 
without any calculation since the remaining degeneracy is a number 
of configurations in which any two {different}  
fields, $\Theta_{\alpha}$ and $\Theta_{\alpha'}$ with $\alpha\neq \alpha'$, 
are placed in the same minimum of the external potential.
In other words, in the bosonized treatment the 
effect of fermionic {exchange} manifests itself 
in a stronger repulsion between the solitons of the same flavor.
This is in agreement with the numerical minimization performed 
in Sec.~\ref{sec:sol4}, Fig.~\ref{fig:sol4}.
The symmetry of the obtained classical ground state is schematically 
illustrated in Fig.~\ref{fig:soliton-sketch}(e).

The change of variables (\ref{transf}) in the potential (\ref{bar-V-12})
yields Eq.~(\ref{bar-L-12}).



\begin{thebibliography}{99}
\bibitem{FSA}
To be precise, this statement is true in 
the forward scattering approximation that holds 
down to undetectably small energy scale, below which further interaction-induced gaps
may open even in the absence of the bare gap, 
$\Delta_0=0$.\cite{Krotov'97,Egger'97,Kane'97,Odintsov'99}
\bibitem{log}
Here we do not distinguish between $\ln(l_s/a)$ and $\ln(l_s/a')$
in the Coulomb formfactor $V(q)$, with $K\gg 1$ implying 
large screening length $l_s\sim \lch \gg a, a'$.
This corresponds to treating the tube capacitance with logarithmic 
accuracy.
\end{thebibliography}
\end{document}